\documentclass{amsproc}

\usepackage{graphicx}% Include figure files
\usepackage{dcolumn}% Align table columns on decimal point
\usepackage{bm}% bold math
\usepackage{tensor}
\usepackage{cancel}
\usepackage{array,amscd,amsmath,amssymb,amsthm,graphicx,mathtools,placeins,slashed,verbatim,tensor,dsfont,enumitem}
\usepackage[bookmarksnumbered,bookmarksopen=true,bookmarksopenlevel=1,pdfstartview=FitH,hidelinks]{hyperref}
\usepackage[all]{hypcap}
\usepackage{rotating}
\usepackage[caption=false]{subfig}
\usepackage{eepic}
\usepackage{braket}
\usepackage{mathrsfs}

\newcommand{\R}{\mathds{R}}
\newcommand{\C}{\mathds{C}}
\newcommand{\Q}{\mathds{H}}
\newcommand{\Oct}{\mathds{O}}
\newcommand{\Al}{\mathds{A}}
\newcommand{\id}{\mathds{1}}

\newcommand{\charge}{\mathcal{Q}}

\newcommand{\rep}[1]{\mathbf{#1}}
\newcommand{\brep}[1]{\mathbf{\overline{#1}}}
\newcommand{\lie}[1]{\mathfrak{#1}}

\newcommand{\cir}[1]{\textcircled{\raisebox{-0.7pt}{#1}}}

\newcommand{\lieh}{\mathfrak{h}}
\newcommand{\lieg}{\mathfrak{g}}

\newcolumntype{L}[1]{>{\raggedright\let\newline\\\arraybackslash\hspace{0pt}}m{#1}}
\newcolumntype{C}[1]{>{\centering\let\newline\\\arraybackslash\hspace{0pt}}m{#1}}
\newcolumntype{R}[1]{>{\raggedLeft\let\newline\\\arraybackslash\hspace{0pt}}m{#1}}

\newcommand{\be}{\begin{equation}}
\newcommand{\ee}{\end{equation}}
\newcommand{\susy}{\mathcal{N}}

\newcommand{\aref}[1]{\hyperref[#1]{appendix~\ref{#1}}}

\theoremstyle{definition}

\theoremstyle{remark}

\numberwithin{equation}{section}

\begin{document}
~~\phantom{hello}   \\

\vspace{-4cm} 
\hfill \hspace{2cm}  {\footnotesize NORDITA 2017-128, DIAS-STP-17-14}\\
\vspace{-1cm} 
\hfill \hspace{2cm}  {\footnotesize IMPERIAL-TP-2017-MJD-02, DFPD/2017/TH-15}\\

\vspace{1cm} 

\vskip 1.5cm

\title{The Mile High Magic Pyramid$^*$}

\author{A.~Anastasiou}
\address{Nordita, KTH Royal Institute of Technology and Stockholm University, Roslagstullsbacken 23, 10691 Stockholm, Sweden}
\email{alexandros.anastasiou@su.se}

\author{L.~Borsten}
%    Address of record for the research reported here
\address{School of Theoretical Physics, Dublin Institute for Advanced Studies,
10 Burlington Road, Dublin 4, Ireland}
%    Current address
\email{leron@stp.dias.ie}
%    \thanks will become a 1st page footnote.
\thanks{The work of LB is supported by a Schr\"odinger Fellowship. SN is supported by a Leverhulme Research Project Grant}

\author{M.~J.~Duff}
%    Address of record for the research reported here
\address{Theoretical Physics, Blackett Laboratory, Imperial College London,
London SW7 2AZ, United Kingdom and Mathematical Institute, University of Oxford, Andrew Wiles Building, Woodstock Road, Radcliffe Observatory Quarter,
Oxford, OX2 6GG, United Kingdom}
\email{m.duff@imperial.ac.uk}
\thanks{MJD is grateful to the Leverhulme Trust for an Emeritus Fellowship and to Philip Candelas for hospitality at the Mathematical Institute, Oxford.  This work was  supported in part by the STFC under rolling grant ST/P000762/1. \\$^*$Lecture delivered by MJD at the 4th Mile High Conference on Nonassociative Mathematics
University of Denver, Denver, Colorado, USA, July 29-August 5, 2017}

\author{A.~Marrani}
%    Address of record for the research reported here
\address{Museo Storico della Fisica e Centro Studi e Ricerche ``Enrico Fermi'',
Via Panisperna 89A, I-00184, Roma, Italy and
Dipartimento di Fisica e Astronomia ``Galileo Galilei'', Universit\`a di Padova, and INFN, sezione di
Padova,
Via Marzolo 8, I-35131 Padova, Italy}
\email{jazzphyzz@gmail.com}
%    \thanks will become a 1st page footnote.

\author{S.~Nagy}
%    Address of record for the research reported here
\address{Centre for Astronomy \& Particle Theory,
University Park,
Nottingham,
NG7 2RD,
United Kingdom}
%    Current address
\email{silvia.nagy1@nottingham.ac.uk}
%    \thanks will become a 1st page footnote.

%    Information for second author
\author{M.~Zoccali}
\address{Theoretical Physics, Blackett Laboratory, Imperial College London,
London SW7 2AZ, United Kingdom}
\email{m.zoccali14@imperial.ac.uk}

%    General info
\date{January 1, 1994 and, in revised form, June 22, 1994.}


\begin{abstract}
Using a unified  formulation of $\mathcal{N} = 1, 2, 4, 8$, super Yang-Mills theories in  $D = 3$ spacetime dimensions with fields valued respectively in $\mathds{R, C, H, O}$, it was shown that tensoring left and right multiplets yields a Freudenthal  magic square of $D = 3$ supergravities. When  tied in with the more familiar $\mathds{R, C, H, O}$ description of super Yang-Mills in $D = 3, 4, 6, 10$ this results in a magic pyramid of supergravities:  the known $4 \times 4$ magic square at the base in $D=3$, a $3\times 3$ square in $D=4$, a $2 \times 2$ square in $D=6$ and Type II supergravity at the apex in $D=10$.
\end{abstract}

\maketitle

\section{Introduction}

In this contribution we describe two very  different ``products'', defined on two  ostensibly unrelated classes of objects, which live in two seemingly disconnected worlds. Nonetheless, they will be shown to meet in an unexpected display of unity. 

The first is the idea that gravity is, in certain regards,  the product of two Yang-Mills gauge theories: ``Gravity$=$Gauge$\times$Gauge''. This is clearly a radical proposal, yet there is growing body of supportive evidence in a range of contexts, from scattering amplitudes \cite{Bern:2008qj, Bern:2010ue, Bern:2010yg} to black hole solutions \cite{Monteiro:2014cda, Cardoso:2016amd}. While the ultimate significance of this program remains to be seen, the proliferation of surprising and illuminating insights uncovered thus far, such as unanticipated cancellations in perturbative quantum gravity \cite{Bern:2014sna},   compels further serious consideration. 

One such surprise brings us to our second incarnation of ``product'',  this time arising in a purely mathematical context. The three associative  normed division algebras, $\mathds{R, C, H}$, provide a concise unified geometric picture of the classical simple Lie algebras, $\mathfrak{so}(N), \mathfrak{su}(N), \mathfrak{usp}(2N)$,   as the isometry algebras of real, complex and quaternionic projective spaces. The final, non-associative, normed division algebra, $\mathds{O}$, does the same for the exceptional Lie algebra   $\mathfrak{f}_4$; it constitutes the isometry algebra of the octonionic projective plane. To proceed further we are required to consider the tensor product of pairs of normed division algebras, $\Al \otimes \tilde{\Al}$, for $\Al,  \tilde{\Al}= \mathds{R, C, H, O}$. By doing so we arrive at the Freudenthal-Rosenfeld-Tits magic square \cite{Freudenthal:1954,Tits:1955, Freudenthal:1959,Rosenfeld:1956, Tits:1966}, a symmetric $4\times 4$ array $\mathfrak{M}(\Al, \tilde{\Al})$ of semi-simple Lie algebras, with the exceptional  algebras $\mathfrak{f}_4, \mathfrak{e}_6, \mathfrak{e}_7, \mathfrak{e}_8$ given by the four  octonionic  entries, $\Al \otimes \mathds{O}$ and $ \mathds{O}  \otimes \tilde{\Al}$.

Through the relationship between supersymmetry and normed division algebras, we shall see that this mathematical magic square is reproduced precisely by the square of supergravity theories generated  by the product of $\mathcal{N} = 1, 2, 4, 8$ super Yang-Mills theories in  $D = 3$ spacetime dimensions \cite{Borsten:2013bp}. Our physical and mathematical magic squares are one and the same! Through the relationship between spacetime symmetries and normed division algebras, this is seen to be  but the base of a \emph{magic pyramid} of Lie algebras and, equivalently, supergravity theories \cite{Anastasiou:2013hba}.

\section{Physics background}

We begin by introducing the physics background necessary to understand the subsequent material. The purpose is to present the basic principles of supersymmetry and use them to sketch the classification of super Yang-Mills and supergravity multiplets with particular emphasis on their global internal symmetries. Then, we introduce the idea of formulating supergravity theories as products of (possibly distinct) pairs of super Yang-Mills theories. We use an explicit example to demonstrate how the symmetries of the factors combine to form those of  the corresponding supergravity. As a motivation for the next section, we close with an observation relating  gravity as the square of Yang-Mills  to the  mathematical   magic square.

\subsection{Supersymmetry}

 Supersymmetry \cite{Golfand:1971iw, Ramond:1971gb, Volkov:1972jx, WESS197439} exchanges  bosons and fermions and, consequently, supersymmetric theories have an equal number of bosonic and fermionic degrees of freedom. In particular,  minimal super Yang-Mills theory (sYM)   has  supersymmetry transformations given schematically by,
\be
\delta A \sim \bar{\epsilon} \psi,\quad  \delta \psi \sim \slashed{F}\epsilon,
\ee 
 where $A$ is the local gauge potential, $F=dA+A\wedge A$, $\slashed{X}=\gamma^{\mu_1\cdots\mu_p}X_{\mu_1\cdots\mu_p}$ and $\psi$ is the fermionic superpartner to $A$, typically referred to as a gluino.  Note, the fermionic supersymmetry transformation parameter, $\epsilon$, is  global for sYM, whereas for supergravity it is local,
 \be
\delta g_{\mu\nu} \sim \bar{\epsilon}(x) \gamma_{(\mu}\Psi_{\nu)},\quad  \delta \Psi_{\mu} \sim \partial_\mu\epsilon(x)+\cdots.
 \ee
 Here, $g_{\mu\nu}$ is the graviton (metric) and  $\Psi_{\mu}$ its fermionic superpartner, the gravitino. 
 For a textbook introduction to supersymmetry and supergravity see \cite{Wess:1992}.  For a more mathematical treatment see \cite{Varadarajan:2004}.

The supersymmetry transformations are induced by fermionic generators, $Q$, which obey (anti)-commutation relations and enhance the spacetime Poincar\'{e} algebra as summarised in \autoref{1}.
\begin{table}[h!]
\centering
\footnotesize
\begin{tabular}{l|ll}
\hline\hline
\\
~~Poincar\'{e}~~ & ~~Supersymmetry~~ & ~~$R$-symmetry~~ 
\\ \\
\hline \\
~~$[M_{[\mu\nu]},M_{[\rho\sigma]}]=\eta_{\nu\rho}M_{[\mu\sigma]}-\eta_{\mu\rho}M_{[\nu\sigma]}-\rho\leftrightarrow\sigma$~~
& ~~$\{Q_{\alpha{I}},Q_\beta^M\}=-\frac{1}{2}\delta_{I}^{M}(\gamma_\mu)_{\alpha\beta}P^\mu+(\dots)$~~
& ~~$[T^{(R)},Q_\alpha^I]=-(R)^{I}{}_{M}Q_\alpha^M$~~
\\ \\
~~$[P_\mu,M_{[\nu\rho]}]=P_\rho\eta_{\mu\nu}-P_\nu\eta_{\rho\mu}$~~ & ~~$[M_{[\mu\nu]},Q_\alpha^I]=\frac{1}{2}(\gamma_{\mu\nu})_{\alpha}{}^{\beta}Q_{\beta}^I$~~
& ~~$[T^{(R)},M_{[\mu\nu]}]=0$
\\ \\
~~$[P_\mu,P_\nu]=0$~~ & ~~$[P_\mu,Q_\alpha^I]=0$~~ & ~~$[T^{(R)},P_\nu]=0$
\\ \\
\hline\hline
\end{tabular}
\caption{The super-Poincar\'{e} Lie algebra.}\label{1}
\end{table}
The number of independent supersymmetry generators is indicated by the index $I=1,\dots,\mathcal{N}$ parametrising the defining representation of the derivations of the supersymmetry algebra, referred to as $R$-symmetry, commuting with the Poincar\'e algebra. 

It is possible to classify these algebras according to the spacetime dimension $D$; to do so, one has to study the associated Clifford algebras and find the set of Lorentz invariant constraints that can be simultaneously imposed on the spinors. The results follow the well known Bott periodic pattern and are summarised in \autoref{2}.
\begin{table}[h!]
\centering
\footnotesize
\begin{tabular}{llccccc}
\hline\hline
\\
~~$D$~~ & ~~Minimal spinor~~ & ~~Reality property ($\mathds{D}$)~~ & ~~Real d.o.f.~~ & ~~$\mathfrak{so}(D-2)$~~ & ~~Representation~~ & ~~$R$-symmetry~~
\\ \\
\hline \\
~~$11$~~ & ~~\textit{Majorana (odd)}~~ & $\mathds{R}$ & ~~32~~ & ~~$\mathfrak{so}(9)$~~ & ~~$\rep{16}$~~ & ~~$\mathfrak{so}(\susy)$~~
\\ \\
~~$10$~~ & ~~\textit{Majorana Weyl}~~ & $\mathds{R}\oplus\mathds{R}$ & ~~16~~ & ~~$\mathfrak{so}(8)$~~ & ~~$\rep{8}_s$ and $\rep{8}_c$~~ & ~~$\mathfrak{so}(\susy)\oplus{\mathfrak{so}}(\tilde\susy)$~~
\\ \\
~~$9$~~ & ~~\textit{Majorana (odd)}~~ & $\mathds{R}$ & ~~16~~ & ~~$\mathfrak{so}(7)$~~ & ~~$\rep{8}$~~ & ~~$\mathfrak{so}(\susy)$~~
\\ \\
~~$8$~~ & ~~\textit{Majorana (even)}~~ & $\mathds{C}$ & ~~16~~ & ~~$\mathfrak{su}(4)$~~ & ~~$\rep{4}\oplus\brep{4}$~~ & ~~$\mathfrak{u}(\susy)$~~
\\ \\
~~$7$~~ & ~~\textit{Symplectic}~~ & $\mathds{H}$ & ~~16~~ & ~~$\mathfrak{usp}(4)$~~ & ~~$\rep{4}$~~ & ~~$\mathfrak{usp}(2\susy)$~~
\\ \\
~~$6$~~ & ~~\textit{Symplectic Weyl}~~ & $\mathds{H}\oplus\mathds{H}$ & ~~8~~ & ~~$2\mathfrak{usp}(2)$~~ & ~~$\rep{(2,1)}$ and $\rep{(1,2)}$~~ & ~~$\mathfrak{usp}(2\susy)\oplus\mathfrak{usp}(2\tilde\susy)$~~
\\ \\
~~$5$~~ & ~~\textit{Symplectic}~~ & $\mathds{H}$ & ~~8~~ & ~~$\mathfrak{usp}(2)$~~ & ~~$\rep{2}$~~ & ~~$\mathfrak{usp}(2\susy)$~~
\\ \\
~~$4$~~ & ~~\textit{Majorana (even)}~~ & $\mathds{C}$ & ~~4~~ & ~~$\mathfrak{u}(1)$~~ & ~~$(+1/2)\oplus(-1/2)$~~ & ~~$\mathfrak{u}(\susy)$~~
\\ \\
~~$3$~~ & ~~\textit{Majorana (odd)}~~ & $\mathds{R}$ & ~~2~~ & ~~$\id$~~ & ~~$1$~~ & ~~$\mathfrak{so}(\susy)$~~
\\ \\
\hline\hline
\end{tabular}
\caption{Minimal spinors and $R$-symmetry algebras. The total number of d.o.f. carried by the supersymmetry generators is called the number of supercharges. It is indicated by $\mathcal{Q}$ and it is the product of $\mathcal{N}$ and the real d.o.f. (cf. 4th column) carried by a single minimal spinor (cf. 3rd column). Note, we associate to each dimension a (direct sum of) division algebra(s) denoted $\mathds{D}$.} 
%Note, for $D=5,6,7$ we regard the sympletic space uniformly as part of the R-symmetry and hence $\mathcal{N}\geq2$ in these cases.}
\label{2}
\end{table}

\begin{table}[h!]
\centering
\footnotesize
\begin{tabular}{C{1.5cm} | C{0.5cm} L{3cm} L{3cm} L {3cm} L{3cm}}
\hline\hline
\\
$D$~~ && $\charge=16$ & $\charge=8$ & $\charge=4$ & $\charge=2$
\\ \\
\hline \\
$10$ && $\id$
\\
&& $\rep{V}_{(1,0)}$
\\ \\
$9$ && $\id$
\\
&& $\rep{V}_{1}$
\\ \\
$8$ && $\mathfrak{u}(1)$
\\
&& $\rep{V}_{1}$
\\ \\
$7$ && $\mathfrak{usp}(2)$
\\
&& $\rep{V}_{1}$
\\ \\
$6$ && $2\mathfrak{usp}(2)$ or $\mathfrak{usp}(4)$ & $\mathfrak{usp}(2)$
\\
&& $\rep{V}_{(1,1)}$ ~~~or $\rep{T}_{(2,0)}$ & $\rep{V}_{(1,0)}$ or $\rep{T}_{(1,0)}$
\\ \\
$5$ && $\mathfrak{usp}(4)$ & $\mathfrak{usp}(2)$
\\
&& $\rep{V}_{2}$ &  $\rep{V}_{1}$
\\ \\
$4$ && $\mathfrak{su}(4)$ & $\mathfrak{u}(2)$ & $\mathfrak{u}(1)$
\\
&& $\rep{V}_{4}$ &  $\rep{V}_{2}$ & $\rep{V}_{1}$
\\ \\
$3$ && $\mathfrak{so}(8)$ & $3\mathfrak{so}(3)$ & $2\mathfrak{so}(2)$ & $\id$
\\
&& $\rep{V}_8$ &  $\rep{V}_4$ & $\rep{V}_2$ &  $\rep{V}_1$
\\ \\
\hline\hline
\end{tabular}
\caption{The super-Yang-Mills theories in $3\leq D\leq10$. We use the notation $\rep{V}_\mathcal{N}$ to denote a vector multiplet with $\mathcal{N}$ supersymmetries. The theories are labelled by their content as well as by their global internal symmetry algebra $\mathfrak{int}$. Theories belonging to the same vertical line can be obtained by toroidal compactification.} 
%The absence of $\mathcal{N}=1$ theories in $D=5,6,7$ reflects the (non-degenerate) symplectic nature of the associated $R$-symmetries.}
\label{3}
\end{table}

With these tools in hand we can sketch how one can classify all possible pure super Yang-Mills multiplets. For a detailed account of this procedure for all supermultiplets in spacetime dimensions $D\leq 11$ see \cite{Strathdee:1986jr}.  For example, in $D=4$, the supercharges are used to construct creation/annihilation  operators,  which raise/lower the helicity of the state they act on. One can build the full spectrum of a  super Yang-Mills multiplet by starting with the $(-1)$ helicity vector and acting repeatedly with a supersymmetry raising operator. The anti-commutation nature of the supersymmetry algebra restricts the number of times one can act with the operators to $\mathcal{N}$. If the highest state of helicity $(+1)$ is not reached then the multiplet is not CPT complete (for a reference, see \cite{Streater:2000}), and the CPT conjugate states must be added in. Note, however, that since we have an upper bound of helicity $(+1)$ this leads to an upper bound on $\mathcal{N}$ giving $\mathcal{N}=4$ as the maximal super Yang-Mills theory in $D=4$. With the $D=4$ classification in hand one can use dimensional reduction/oxidation to classify all super Yang-Mills theories in $3\leq D\leq10$ as summarised in \autoref{3}.

Note that the Yang-Mills action functional is invariant under an internal\footnote{In the sense that it commutes with the Poincar\'e algebra.} symmetry algebra $\mathfrak{int}$, which is identical to the $R$-symmetry algebra, except in the following cases:
\begin{itemize}
\item In $D=4$ the maximal theory is CPT self-conjugate because the raising procedure builds the full multiplet without the need of adding the CPT conjugate states. A consequence of this is that the $\mathfrak{u}(1)$ part of the $R$-symmetry cannot be supported.
\item Since we are working with massless on-shell states, vectors dualise to scalars in $D=3$. As a result, in the $\mathcal{N}=2,4$ there is an enhancement of $R$ to the full $\mathfrak{int}$ as indicated in the last row of \autoref{3}.
\end{itemize}

One can follow the same raising procedure for a lowest helicity state $(-2)$ corresponding to that of a graviton and obtain the spectrum of a supergravity multiplet. It is simple to see that the maximal multiplet (in the sense that we restrict to helicity $\leq 2$) corresponds to $Q=32$. For example, this leads to $\susy=8$ supergravity in $D=4$ and the unique $\susy=1$ supergravity in $D=11$. For $\susy>$ half-maximal there is a unique supergravity, but in general the classification of  supergravity theories is rather involved  and the details of this are well beyond the scope of this article; for a comprehensive account see \cite{Salam:1989fm}. What we are interested in here are the symmetries of the supergravity theories and this is what we will briefly discuss now. The scalars of a supergravity theory parametrise a target space called the scalar manifold, which in all cases of interest here will be a Riemannian symmetric space $\mathcal{G}/\mathcal{H}$, for $\mathcal{G}$ a non-compact semi-simple Lie group and $\mathcal{H}$ its maximal compact subgroup. In the context of supergravity, $\mathcal{G}$ corresponds to the global internal symmetry group of the theory, also referred to in the physics literature as the U-duality group due to its relation to the U-dualities of M-theory, while $\mathcal{H}$ corresponds to its maximal compact subgroup. Alternatively, $\mathcal{H}$ can be thought of as the largest subgroup of $\mathcal{G}$ linearly realised on all fields of the supergravity theory. For a review on symmetric scalar manifolds in this framework, see \cite{Ferrara:2008de}.

\subsection{Gravity as the square of Yang-Mills}

The idea that gravity, in certain respects, can be re-conceived as the product of two Yang-Mills gauge theories has undergone a renaissance over the last decade, initiated  by the remarkable Bern-Carrasco-Johansson double-copy amplitude construction \cite{Bern:2008qj, Bern:2009kd, Bern:2010ue, Bern:2010yg, Bern:2012cd, Bern:2013uka,  Bern:2013yya, Bern:2013qca,Bern:2014lha, Bern:2014sna}.  There is now an extensive array of gravitational theories and phenomena admitting a Yang-Mills squared origin. See for example  \cite{Monteiro:2011pc, Carrasco:2012ca,  Huang:2012wr, Bargheer:2012gv, Monteiro:2013rya, Monteiro:2014cda,  Anastasiou:2014qba, Nagy:2014jza, Johansson:2014zca, Chiodaroli:2014xia, Chiodaroli:2015rdg, Chiodaroli:2015wal, Anastasiou:2015vba, Borsten:2015pla, Luna:2015paa, Luna:2016due, Luna:2016hge, White:2016jzc, Goldberger:2016iau, Cardoso:2016ngt, Cardoso:2016amd,   Chiodaroli:2016jqw, Johansson:2017bfl,  Goldberger:2017frp, Anastasiou:2017nsz, Bahjat-Abbas:2017htu,  Carrillo-Gonzalez:2017iyj, DeSmet:2017rve, Chiodaroli:2017ehv, Luna:2017dtq} and the references therein.

The basic intuition is conveyed by the illustrative ``product'':
\begin{equation}\label{intuition}
A_\mu(x)\; \text{``$\otimes$''\;}  \tilde{A}_\nu(x) = g_{\mu\nu}(x) + \cdots
\end{equation}
Here, $A_\mu$ and $\tilde{A}_\nu$ are the gauge potentials of two distinct Yang-Mills theories, which we will refer to as \emph{Left} and \emph{Right}, respectively. From now on, all fields in the \textit{Right} theory will be denoted with a tilde to distinguish them from those in the \textit{Left} theory. In fact, the product ``$\otimes$'' appearing in  \eqref{intuition} can be made precise for arbitrary spacetime fields, belonging to a very general class of gauge theories, to linear approximation \cite{Anastasiou:2014qba}. However, here it will suffice to confine our attention to the conventional tensor product of the corresponding on-shell massless states, which carry representations of the Lie algebra $\mathfrak{so}(D-2)$ of the spacetime little group, as well as the internal global symmetry algebra $\mathfrak{int}$ and the internal local gauge algebra.

 Upon squaring, the \emph{Left} and \emph{Right} spacetime little group algebras are identified, while the internal algebras remain independent. Hence, the  product states are $(\mathfrak{so}(D-2)\oplus\mathfrak{int} \oplus \mathfrak{\tilde{int}})$-modules.  In particular, the on-shell  metric corresponds to  a rank-$2$ symmetric traceless tensor of $\mathfrak{so}(D-2)$ originating from the tensor product of two $\mathfrak{so}(D-2)$ vectors, which correspond to the on-shell states of the \emph{Left} and \emph{Right}  gauge potentials (which are always singlets under $\mathfrak{int}$ and  $\mathfrak{\tilde{int}}$).  It follows that   $\mathfrak{int} \oplus \mathfrak{\tilde{int}}$ constitutes a sub-algebra of $\mathfrak{h}$, where $\mathfrak{h}$ is the Lie algebra of $\mathcal{H}$. As explained in \cite{Anastasiou:2013hba, Anastasiou:2015vba, Anastasiou:2017nsz}, $\mathfrak{int} \oplus \mathfrak{\tilde{int}}$ gets enhanced to the full $\mathfrak{h}$, where the extra generators are schematically given by the tensor product of the \emph{Left} and \emph{Right} supersymmetry generators, $Q\otimes\tilde{Q}$.

Of course, the resulting gravitational states must be scalars
under the gauge groups $G\times\tilde{G}$. In order to enforce this, we conjecture \cite{Borsten:2013bp} that the  product involves, apart from the \textit{Left} and \textit{Right} super-Yang-Mills, another field which we call the bi-adjoint \textit{spectator scalar}, $\Phi$. Note, the spectator field has proven crucial in a number of related, yet distinct, realisations of the ``gravity = gauge $\times$ gauge'' paradigm \cite{Hodges:2011wm, Cachazo:2013iea, Monteiro:2013rya, Luna:2016hge}. In this context, $\Phi$ allows for arbitrary and independent   $G$ and $\tilde{G}$ at the level of spacetime fields. Moreover, it turns out to be essential for reproducing the local symmetries of (super)gravity from those of the two (super) Yang-Mills factors to linear order \cite{Anastasiou:2014qba, Borsten:2015pla, Anastasiou:2016csv, Anastasiou:2017nsz} as well as ensuring that the functional degrees of freedom can be correctly matched. 

Let us put these ideas into practice through the canonical $D=4$ example of $\susy=8$ supergravity as the product of two $\susy=4$ super Yang-Mills theories. The states  of the \emph{Left} (\emph{Right}) multiplet are labelled by their helicity charges $q$ $(\tilde{q})$ under $\mathfrak{so}(2)\cong\mathfrak{u}(1)_{st}$ and their representation under the internal  $\mathfrak{int}\cong\mathfrak{su}(4)$ ($\tilde{\mathfrak{int}}\cong\mathfrak{su}(4)$). According to the discussion above, the resulting gravity fields should carry representations under:
\be
\mathfrak{u}(1)_{st}\oplus \mathfrak{su}(4)\oplus \mathfrak{su}(4)\oplus \mathfrak{u}(1)_d,\text{~~~~where~~~~}q_{st}=q+\tilde{q}\text{~~~~and~~~~}q_d=q-\tilde{q}.
\ee
The $\mathfrak{u}(1)_d$ factor above, which is special to the case of $D=4$, is observed to be a required symmetry of the gravitational theory in the case of symmetric scalar manifolds \cite{Anastasiou:2013hba, Chiodaroli:2015wal, Anastasiou:2017nsz}.

Using these rules we perform the squaring calculation\footnote{Here and henceforth we double the $\mathfrak{u}(1)_{st}$ charge, for notational convenience.} as in \autoref{N=4}.
\begin{table}[h!]
\centering
\footnotesize
\begin{tabular}{C{2cm}|C{5cm} C{5cm} C{3cm}}
\hline\hline
\\
$\rep{V}_4\backslash\rep{\tilde{V}}_4$ & $\rep{1}_{-2}+\rep{1}_{2}$
& $\rep{4}_{-1}+\brep{4}_{1}$ & $\rep{6}_{0}$ \\
\\
\hline
\\
$\rep{1}_{-2}+\rep{1}_{2}$ & $\rep{(1,1)}^0_{4}+\rep{(1,1)}^0_{-4}+\rep{(1,1)}^4_{0}+\rep{(1,1)}^{-4}_{0}$
& $\rep{(1,4)}^3_{1}+(\rep{1},\brep{4})^{-3}_{-1}+\rep{(1,4)}^{-1}_{-3}+(\rep{1},\brep{4})^1_{3}$ & $\rep{(1,6)}_2^{2}+\rep{(1,6)}_{-2}^{-2}$ \\
\\
$\rep{4}_{-1}+\brep{4}_{1}$ & $\rep{(4,1)}^{-3}_{1}+(\brep{4},\rep{1})^{3}_{-1}+\rep{(4,1)}^{1}_{-3}+(\brep{4},\rep{1})^{-1}_{3}$ & $\rep{(4,4)}^{0}_{-2}+(\brep{4},\brep{4})^{0}_{2}+(\rep{4},\brep{4})^{-2}_{0}+(\brep{4},\rep{4})^{2}_{0}$
& $\rep{(4,6)}^{-1}_{-1}+(\brep{4},\rep{6})^{1}_{1}$ \\
\\
$\rep{6}_{0}$ & $\rep{(6,1)}^{-2}_{2}+\rep{(6,1)}^{2}_{-2}$
& $\rep{(6,4)}^{1}_{-1}+(\rep{6},\brep{4})^{-1}_{1}$ & $\rep{(6,6)}^0_{0}$
\\
\\
\hline
\hline
\end{tabular}
\caption{Squaring table for $\rep{V}_4\otimes\rep{\tilde{V}}_4$ in $D=4$. The pairs $(\mathbf{x}, \tilde{\mathbf{x}})$ denote the $\mathfrak{int}\oplus\tilde{\mathfrak{int}}\cong \mathfrak{su}(4)\oplus \mathfrak{su}(4)$ representations carried by the supergravity states. The subscripts (superscripts) denote the $\mathfrak{u}(1)_{st}$ ($\mathfrak{u}(1)_{d}$) charge, $q_{st}$ ($q_d)$,  carried by the states.}\label{N=4}
\end{table}
By collecting states of a given  helicity we see that there is an obvious enhancement of $\mathfrak{su}(4)\oplus \mathfrak{su}(4)\oplus \mathfrak{u}(1)_d\supset \mathfrak{su}(8)$ such that each helicity state carries the correct $\mathcal{N}=8$ supergravity representation:
\begin{align}
&\text{Graviton: }&&\mathbf{(1,1)}_0&&\rightarrow&&\rep{1}\\
&\text{Gravitini: }&&\mathbf{(4,1)}_{1}+\mathbf{(1,4)}_{-1}&&\rightarrow&&\rep{8}\\
&\text{$1$-form: }&&\mathbf{(6,1)}_{2}+\mathbf{(1,6)}_{-2}+\mathbf{(4,4)}_{0}&&\rightarrow&&\rep{28}\\
&\text{Spinors: }&&\mathbf{(\bar{4},1)}_{3}+\mathbf{(1,\bar{4})}_{-3}+\mathbf{(6,4)}_{1}+\mathbf{(4,6)}_{-1}&&\rightarrow&&\rep{56}\\
&\text{Scalars: }&&\mathbf{(1,1)}_4+\mathbf{(1,1)}_{-4}+\mathbf{(\bar{4},4)}_{2}+\mathbf{(4,\bar{4})}_{-2}+\mathbf{(6,6)}_{0}&&\rightarrow&&\rep{70}
\end{align}

Since the scalars transform irreducibly under $\mathfrak{su}(8)$, the scalar manifold is necessarily locally homogeneous \cite{ArkaniHamed:2008gz, Chiodaroli:2015wal}. Moreover, as explained above, the presence of the global $\mathfrak{u}(1)_d$ implies that the coset is symmetric. The unique non-compact candidate for $\mathfrak{g}$ is then $\mathfrak{e}_{7(7)}$, the global symmetry algebra of $\susy = 8$ supergravity.

At this point one could repeat this exercise for every possible pair $(\susy,\tilde\susy)$ in each dimension $D$ and build the scalar coset of the associated supergravity. Here, we instead seek a single formula that takes $(D,\mathcal{Q},\tilde\charge)$ as input and gives the resulting $\mathfrak{g}$ as output.

 A case-by-case study of all the possible super Yang-Mills products in $D=3$ results in the  supergravity theories given in \autoref{d3sugras}.
\begin{table}[h!]
\centering
\footnotesize
\begin{tabular}{C{1.5cm} | C{0.25cm} L{3cm} L{3cm} L{3cm} L{3cm}}
\hline\hline
\\
$\mathcal{N}~\backslash~\tilde{\mathcal{N}}$ && $8$ & $4$ & $2$ & $1$
\\ \\
\hline \\
$8$ && $\lieg=\mathfrak{e}_{8(8)}$ & $\lieg=\mathfrak{e}_{7(-5)}$ & $\lieg=\mathfrak{e}_{6(-14)}$
& $\lieg=\mathfrak{f}_{4(-20)}$
\\
&& $\lieh=\mathfrak{so}(16)$ & $\lieh=\mathfrak{so}(12)\oplus\mathfrak{so}(3)$
& $\lieh=\mathfrak{so}(10)\oplus\mathfrak{so}(2)$ & $\lieh=\mathfrak{so}(9)$ 
\\ \\
$4$ && $\lieg=\mathfrak{e}_{7(-5)}$ & $\lieg=\mathfrak{so}(8,4)$ & $\lieg=\mathfrak{su}(4,2)$
& $\lieg=\mathfrak{usp}(4,2)$
\\
&& $\lieh=\mathfrak{so}(12)\oplus\mathfrak{so}(3)$ & $\lieh=\mathfrak{so}(8)\oplus2\mathfrak{so}(3)$
& $\lieh=\mathfrak{so}(6)\oplus\mathfrak{so}(3)\oplus\mathfrak{so}(2)$ & $\lieh=\mathfrak{so}(5)\oplus\mathfrak{so}(3)$
\\ \\
$2$ && $\lieg=\mathfrak{e}_{6(-14)}$ & $\lieg=\mathfrak{su}(4,2)$ & $\lieg=2\mathfrak{su}(2,1)$
& $\lieg=\mathfrak{su}(2,1)$
\\
&& $\lieh=\mathfrak{so}(10)\oplus\mathfrak{so}(2)$ & $\lieh=\mathfrak{so}(6)\oplus\mathfrak{so}(3)\oplus\mathfrak{so}(2)$
& $\lieh=\mathfrak{so}(4)\oplus2\mathfrak{so}(2)$ & $\lieh=\mathfrak{so}(3)\oplus\mathfrak{so}(2)$
\\ \\
$1$ && $\lieg=\mathfrak{f}_{4(-20)}$ & $\lieg=\mathfrak{usp}(4,2)$ & $\lieg=\mathfrak{su}(2,1)$
& $\lieg=\mathfrak{so}(2,1)$
\\
&& $\lieh=\mathfrak{so}(9)$ & $\lieh=\mathfrak{so}(5)\oplus\mathfrak{so}(3)$ & $\lieh=\mathfrak{so}(3)\oplus\mathfrak{so}(2)$ & $\lieh=\mathfrak{so}(2)$
\\ \\
\hline\hline
\end{tabular}
\caption{The magic square of supergravities in $D=3$. The rows and columns determine a pair of sYM theories while the table entries correspond to the algebras of $\mathcal{G}$ resp. $\mathcal{H}$ of the (symmetric) scalar coset $\mathcal{G}/\mathcal{H}$ of the resulting supergravity theory.}\label{d3sugras}
\end{table}
This corresponds precisely to the \emph{Lorentzian Freudenthal-Rosenfield-Tits magic square} \cite{Freudenthal:1954, Tits:1955, Rosenfeld:1956} (see also \cite{Vinberg:1966, Tits:1966, Barton:2003, Cacciatori:2012cb}): a $4\times 4$ symmetric array of Lie algebras given in \autoref{LFRT}, whose entries are determined by a pair of normed division  algebras (NDAs) (or more generally a pair of composition algebras) by the formula \cite{Borsten:2013bp, Anastasiou:2013hba},
\be
\mathfrak{M}(\Al, \tilde{\Al}):=\mathfrak{tri}(\Al)\oplus\mathfrak{tri}(\tilde{\Al})+3(\Al\otimes\tilde{\Al}).
\ee
This particular realisation of the magic square formula is adapted from the Barton-Sudbery triality construction \cite{Barton:2003}. The notation used here will be made clear in \autoref{section3}.

\begin{table}[h!]
\centering
\footnotesize
\begin{tabular}{C{1.5cm} | C{0.25cm} L{3cm} L{3cm} L{3cm} L{3cm}}
\hline\hline
\\
$\Al~\backslash~\tilde{\Al}$ && $\Oct$ & $\Q$ & $\C$ & $\R$
\\ \\
\hline \\
$\Oct$ && $\mathfrak{e}_{8(8)}$ & $\mathfrak{e}_{7(-5)}$ & $\mathfrak{e}_{6(-14)}$
& $\mathfrak{f}_{4(-20)}$
\\ \\
$\Q$ && $\mathfrak{e}_{7(-5)}$ & $\mathfrak{so}(8,4)$ & $\mathfrak{su}(4,2)$
& $\mathfrak{usp}(4,2)$
\\ \\
$\C$ && $\mathfrak{e}_{6(-14)}$ & $\mathfrak{su}(4,2)$ & $2\mathfrak{su}(2,1)$
& $\mathfrak{su}(2,1)$
\\ \\
$\R$ && $\mathfrak{f}_{4(-20)}$ & $\mathfrak{usp}(4,2)$ & $\mathfrak{su}(2,1)$
& $\mathfrak{so}(2,1)$
\\ \\
\hline\hline
\end{tabular}
\caption{The Lorentzian FRT magic square of Lie algebras \cite{Elduque:2006, Cacciatori:2012cb, Anastasiou:2013hba}. See also section 12 of \cite{Jacobson:1971}. }\label{LFRT}
\end{table}
Remarkably, the entries of the magic square match exactly those of $\lieg$ in the $D=3$ squaring table. This is the  first instance of the complete magic square with these real forms appearing  in supergravity. The goal of the next section is to study the NDAs and their relation to sYM theories. Then, in the final section we will give a physical interpretation to the magic square formula and generalise it to higher dimensions.

\section{The Normed Division Algebras}\label{section3}

\subsection{General remarks}

As promised we will now go through a lightning introduction to the NDAs and their basic properties. See \cite{Baez:2001dm} for an excellent review. An algebra $\Al$ is a (for us real) vector space equipped with
a bilinear multiplication rule and a unit element. We say $\Al$ is a division
algebra if, given $x,y\in\Al$ with $xy=0$, then either $x=0$ or $y=0$. A
normed division algebra is an algebra $\Al$
equipped with a positive-definite norm satisfying the condition 
\be
|\hspace{-0.2mm}|xy|\hspace{-0.2mm}|=|\hspace{-0.2mm}|x|\hspace{-0.2mm}|\hspace{0.4mm}|\hspace{-0.2mm}|y|\hspace{-0.2mm}|,
\ee
which also implies $\Al$ is a division algebra. From now on it shall be understood
that the term `division algebra' is short for `normed division algebra',
since we shall have no cause to use division algebras that are not normed. A remarkable theorem due to Hurwitz \cite{Hurwitz:1898} states
that there are precisely  four normed division algebras: the real numbers $\R$,
the complex numbers $\C$,
the quaternions $\Q$ and the octonions $\Oct$. The algebras have dimensions
$n=1,2,4$ and $8$, respectively. They can be constructed, one-by-one, using
the Cayley-Dickson doubling method, starting with $\R$; the complex numbers
are pairs of real numbers equipped with a particular multiplication rule,
a quaternion is a pair of complex numbers and an octonion is a pair of quaternions.
At the level of vector spaces:
\be
\begin{split}
\C &\cong\R^2,\\
\Q &\cong\C^2 \hspace{0.4mm} \cong\R^4,\\
\Oct &\cong\Q^2\cong\C^4\cong\R^8.
\end{split}
\ee
The real numbers are ordered, commutative and associative, but with each
doubling one such property is lost: $\C$ is commutative and associative,
$\Q$ is associative, $\Oct$ is \emph{non-associative}: The Cayley-Dickson
procedure yields an infinite sequence of algebras, but in doubling the octonions
to obtain the 16-dimensional `sedenions' $\mathds{S}$ the division algebra property is
lost, as  realised  by the discoverer of the octonions, John T.~Graves \cite{Baez:2001dm}.\\
\begin{table}[h!]
\centering
\footnotesize
\begin{tabular}{L{2.3cm} | C{0.5cm} C{2cm} C{2cm} C{2cm} C{2cm}}
\hline\hline
\\
~~Property && $\R$ & $\C$ & $\Q$ & $\Oct$
\\ \\
\hline \\
~~Ordered && $\checkmark$ & $\times$  & $\times$ & $\times$
\\ \\
~~Commutative && $\checkmark$ & $\checkmark$ & $\times$ & $\times$
\\ \\
~~Associative && $\checkmark$ & $\checkmark$ & $\checkmark$ & $\times$
\\ \\
~~Division && $\checkmark$ & $\checkmark$ & $\checkmark$ & $\checkmark$
\\ \\
\hline\hline
\end{tabular}
\caption{Properties of the NDAs.}
\end{table}

\begin{figure}[h!]
  \centering
    \includegraphics[width=0.5\textwidth]{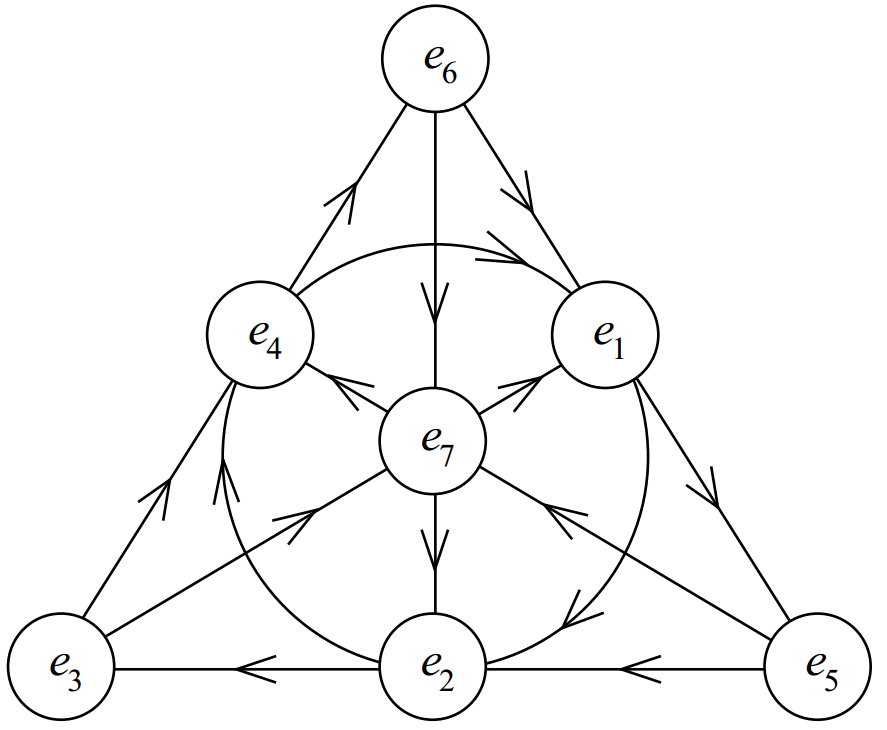}
  \caption[The Fano plane.]{The Fano plane. Each oriented
line corresponds to a quaternionic subalgebra.}\label{FANO}
\end{figure}

On occasion  it will be useful to denote the division algebra of dimension
$n$ by $\Al_n$. A division algebra element $x\in\Al$ is written as the linear combination
of $n$ basis elements with real coefficients: $x=x_{a}e_{a}$, with $x_a\in\R$
and $a=0,\cdots,(n-1)$. The first basis element $e_0=1$ is real, while the
other $(n-1)$ bases $e_i$ are imaginary:
\be
e_0^2=1,~~~~e_i^2=-1,
\ee
where $i=1,\cdots,(n-1)$. In analogy with the complex case, we define a conjugation
operation indicated by *, which changes the sign of the imaginary basis elements:
\be
{e_0}^*=e_0,~~~~{e_i}^*=-e_i.
\ee
It is natural then to define the real and imaginary parts of $x\in\Al$ by
\be
\text{Re}(x):=\frac{1}{2}(x+x^*)=x_0,~~~~~~\text{Im}(x):=\frac{1}{2}(x-x^*)=x_ie_i.
\ee
Note that this differs slightly with the convention typically used for the
complex numbers (since $\text{Im}(x_0+x_1e_1)=x_1e_1$ rather than $x_1$).
The multiplication
rule for the basis elements of a general division algebra is given by:
\begin{align}
&e_a e_b=\Gamma_{ab}^{c}e_c=(+\delta_{a0}\delta_{bc}+\delta_{b0}\delta_{ac}-\delta_{c0}\delta_{ab}+C_{abc})e_c\label{OCTMULT1}\\
&e_a^*e_b=\bar\Gamma_{ab}^{c}e_c=(+\delta_{a0}\delta_{bc}-\delta_{b0}\delta_{ac}+\delta_{c0}\delta_{ab}-C_{abc})e_c\quad\Rightarrow\quad\Gamma^a_{bc}=\bar\Gamma^a_{cb}\label{OCTMULT2}
\end{align}
The tensor $C_{abc}$ is totally antisymmetric with $C_{0ab}=0$, implying it is identically zero for $\Al=\R,\C$. For the quaternions
$C_{ijk}$ is simply the permutation symbol $\varepsilon_{ijk}$, while for
the octonions the non-zero $C_{ijk}$ are specified by the set $\bold{L}$
of oriented lines of the Fano plane, which can be used
as a mnemonic for octonionic multiplication - see Fig. \ref{FANO}:
\be
\begin{split}
C_{ijk}(\Al) &= \begin{cases} 0 &\mbox{for } \Al=\R,\C \\
1 \hspace{0.1cm}\text{ if }\hspace{0.1cm}ijk=123 & \mbox{for } \Al=\Q  \\
1 \hspace{0.1cm}\text{ if }\hspace{0.1cm}ijk\in\bold{L} & \mbox{for } \Al=\Oct,
\end{cases}\\
\hspace{0.1cm}\text{ where }\hspace{0.1cm}\bold{L}&=\{124,235,346,457,561,672,713\}.
\end{split}
\ee
It is useful to remember that adding 1 (modulo 7) to each of the digits labelling
a line in $\bold{L}$ produces the next line. For example, $124\rightarrow235$.\\
Restricting to any single line of the Fano plane restricts the octonions
to a quaternionic subalgebra so that $C_{ijk}$ reduces to the permutation
symbol $\varepsilon_{ijk}$. For example, the subalgebra spanned by $\{e_0,e_4,e_5,e_7\}$
is isomorphic to the quaternions. The complement of a line is called a quadrangle $\bold{Q}$ and all seven of them appear through the tensor $Q_{abcd}$ which is totally antisymmetric with $Q_{0abc}=0$, and
the non-zero $Q_{ijkl}$ are given by:
\be
Q_{ijkl}=1\hspace{0.2cm}\text{if}\hspace{0.2cm}ijkl\in\bold{Q}=\{3567,4671,5712,6123,7234,1345,2456\}.
\ee
Since a quadrangle is the complement of a line in the Fano plane, by definition,
the tensors $Q_{ijkl}$ and $C_{ijk}$ are dual to one another:
\be\label{QC}
Q_{ijkl}=-\frac{1}{3!}\varepsilon_{ijklmnp}C_{mnp}.
\ee
Although octonions are non-associative,
they enjoy the weaker property
of \emph{alternativity}. An algebra $\Al$ is alternative if and only if for
all $x,y\in\Al$ we have:
\be\label{three_cond}
(xx)y=x(xy),~~~~~~(xy)x=x(yx),~~~~~~(yx)x=y(xx)
\ee
This
property is trivially satisfied by the three associative division algebras
$\R,\C$ and $\Q$, and so we conclude that the division algebras are alternative.
The three conditions \eqref{three_cond} can be neatly summed up if we define a trilinear map
called the associator given by:
\be
[x,y,z]=(xy)z-x(yz),~~~~x,y,z\in\Al,
\ee
which measures the failure of associativity. An algebra $\Al$ is then alternative
if and only if the associator is an antisymmetric function of its three arguments. Using \eqref{OCTMULT2} and \eqref{QC} we have
\be
[e_a,e_b,e_c]=2Q_{abcd}e_d.
\ee

\subsection{Relation to Lie algebras}

The division algebras are closely related to Lie algebras in various guises,
with exceptional algebras related to the exceptional status of the octonions
as the only non-associative division algebra. We will show how the division
algebras introduce a natural action of the algebras $\mathfrak{so}(n)\supset \mathfrak{so}(n-1)\supset
\mathfrak{aut}(\Al_n)$ where the latter corresponds to the algebra of derivations of the
division algebra $\Al_n$. 
Once these relations are established we will introduce the so-called triality
algebra $\mathfrak{tri}(\Al_n)$. Then, we
will show how the well known relation between the classical algebras and $\R,\C,\Q$
can be extended to include the octonions. Finally, we will show how all these relations can be brought together to form even larger algebras leading to the magic square construction.

\subsubsection{$\mathfrak{so}(n)$, $n=1, 2, 4, 8$}

These Lie algebras have the remarkable property of having vector, spinor and conjugate spinor representations of the same (real) dimensions, a property best known as triality. For example, in the case of $\mathfrak{so}(8)$ the vector, spinor and conjugate spinor representations, $\rep{8}_v,\rep{8}_s,\rep{8}_c$, can all be realised as $\Oct$ and the action of $\mathfrak{so}(8)$  will take a multiplicative form, different in each case. By determining the correct  action  of $\mathfrak{so}(8)$ on $\Oct$, we obtain all cases when suitably interpreted.

The $\rep{8}_v,\rep{8}_s,\rep{8}_c$ of $\mathfrak{so}(8)$ transform as:
\begin{align}
\delta V_c&=-\frac{1}{2}\theta_{ab}(M_v^{[ab]})_{c}{}^{d}V_d,&&(M_v^{[ab]})_{c}{}^{d}=2\delta_{c}^{[a}\delta^{b]d},\\
\delta\psi_\alpha&=-\frac{1}{2}\theta_{ab}(M_v^{[ab]})_{\alpha}{}^{\beta}\psi_\beta,&&(M_v^{[ab]})_{\alpha}{}^{\beta}=\frac{1}{2}(\Gamma^{[a}\bar\Gamma^{b]})_{\alpha}{}^{\beta},\label{SPINTRA1}\\
\delta\chi_{\dot\alpha}&=-\frac{1}{2}\theta_{ab}(M_v^{[ab]})_{\dot\alpha}{}^{\dot\beta}\chi_{\dot\beta},&&(M_v^{[ab]})_{\dot\alpha}{}^{\dot\beta}=\frac{1}{2}(\bar\Gamma^{[a}\Gamma^{b]})_{\dot\alpha}{}^{\dot\beta},\label{SPINTRA2}
\end{align}
respectively. One can check that in all three cases the indices $a,\alpha,\dot\alpha=0,\dots 7$ can be raised/lowered by a $\delta$ matrix and that, in all three cases, the generators can be chosen to be real antisymmetric matrices. These are consequences of the so-called $\mathfrak{so}(8)$ triality, which implies that we can simply use a single index $a$ for all three representations,  which will be distinguished only by the symbols $V,\psi,\chi$. Furthermore, since the gamma matrices appearing in the octonionic multiplication $(\ref{OCTMULT1})$ and $(\ref{OCTMULT2})$ satisfy the properties in $(\ref{SPINTRA1})$ and $(\ref{SPINTRA2})$, we can parametrise $V=V_ae_a,\psi=\psi_ae_a,\chi=\chi_ae_a$ with transformations:
\begin{align}
\delta V&=-\frac{1}{4}\theta_{ab}\Big((e_ae_b^*)V-V(e_a^*e_b)+[e_a,e_b,V]\Big):=\theta V+V\bar\theta+\hat{a}V\\
\delta\psi&=-\frac{1}{4}\theta_{ab}\Big((e_a^*e_b)\psi+[e_a,e_b,\psi]\Big):=-\bar\theta\psi+\hat{a}\psi\\
\delta\chi&=-\frac{1}{4}\theta_{ab}\Big((e_ae_b^*)\chi+[e_a,e_b,\chi]\Big):=\theta\chi+\hat{a}\chi
\end{align}
At first sight this form might not seem very illuminating, but  the standard transformations follow straightforwardly for all $\mathfrak{so}(n)$ by simply changing the algebra $\Al_n$ as in \autoref{vsc}.
\begin{table}[h!]
\centering
\footnotesize
\begin{tabular}{C{1cm} C{2cm} C{2cm} C{2cm} C{2cm} C{2cm}}
\hline\hline
\\
$\Al$ & $\mathfrak{so}(n)$ & $V$ & $\psi$ & $\chi$ & Parameters
\\ \\
\hline \\
$\Oct$ & $\mathfrak{so}(8)$ & $\rep{8}_v$  & $\rep{8}_s$ & $\rep{8}_c$ & $\hat{a}\neq0$, $\theta\neq\bar\theta$ 
\\ \\
$\Q$ & $2\mathfrak{su}(2)$ & $\rep{(2,2)}$ & $2\times\rep{(2,1)}$ & $2\times\rep{(1,2)}$ & $\hat{a}=0$, $\theta\neq\bar\theta$ 
\\ \\
$\C$ & $\mathfrak{u}(1)$ & $(+2)$ & $(-1)$ & $(+1)$ & $\hat{a}=0$, $\theta=\bar\theta$ 
\\ \\
\hline\hline
\end{tabular}
\caption{Vector, spinor and conjugate spinor representations of $\mathfrak{so}(n)$.}\label{vsc}
\end{table}

\subsubsection{$\mathfrak{so}(n-1)$}

Using the result of the previous section  the vector, spinor and conjugate spinor representations of  $\mathfrak{so}(n-1)$ follow. Once again we start with the octonionic case, derive the transformations and simply generalise them to all $n$ by changing their interpretation. Since  under $\mathfrak{so}(8)\subset \mathfrak{so}(7)$,
\begin{align}
\rep{8}_v&\rightarrow \rep{1}+\rep{7}\\
\rep{8}_s&\rightarrow \rep{8}\\
\rep{8}_c&\rightarrow \rep{8}
\end{align}
it is natural to identify the vector representation with the imaginary subspace $\text{Im}\Oct$,  $V=V_ie_i$,  and set $\theta_{0i}=0$, yielding,
\begin{align}\label{belli}
\delta V&=-\frac{1}{4}\theta_{ij}\Big(-(e_ie_j)V+V(e_ie_j)+[e_i,e_j,V]\Big):=\vartheta V-V\vartheta+\hat{a}V\\\label{belli2}
\delta\psi&=-\frac{1}{4}\theta_{ij}\Big(-(e_ie_j)\psi+[e_i,e_j,\psi]\Big):=\vartheta\psi+\hat{a}\psi\\\label{belli3}
\delta\chi&=-\frac{1}{4}\theta_{ij}\Big(-(e_ie_j)\chi+[e_i,e_j,\chi]\Big):=\vartheta\chi+\hat{a}\chi
\end{align}
As before, the transformations for all $\mathfrak{so}(n)$ follow by simply changing the NDA used as given in \autoref{son-1}.
\begin{table}[h!]
\centering
\footnotesize
\begin{tabular}{C{1cm} C{2cm} C{2cm} C{2cm} C{2cm} C{2cm}}
\hline\hline
\\
$\Al$ & $\mathfrak{so}(n-1)$ & $V$ & $\psi$ & $\chi$ & Parameters
\\ \\
\hline \\
$\Oct$ & $\mathfrak{so}(7)$ & $\rep{7}$  & $\rep{8}$ & $\rep{8}$ & $\hat{a}\neq0$
\\ \\
$\Q$ & $\mathfrak{su}(2)$ & $\rep{3}$ & $\rep{2}$ & $\rep{2}$ & $\hat{a}=0$ 
\\ \\
\hline\hline
\end{tabular}
\caption{Vector, spinor and conjugate spinor representations of $\mathfrak{so}(n-1)$.}\label{son-1}
\end{table}

\subsubsection{$\mathfrak{aut}(\Al)$}

Using the transformations \eqref{belli}, \eqref{belli2} and \eqref{belli3}, it is straightforward to recover the standard  action of $\mathfrak{g}_2$ on the octonions via the algebra of derivations. The algebra of derivations $\mathfrak{aut}(\Al)$ of an algebra $\Al$ is the algebra of linear transformations that preserve multiplication in $\Al$. The derivations preserve the norm and leave the real subspace of the division algebra invariant and thus $\mathfrak{aut}(\Al)$ must be a sub-algebra of $\mathfrak{so}(n-1)$ acting non-trivially only on the imaginary subspace. More specifically, it is the sub-algebra of $\mathfrak{so}(n-1)$ which leaves the structure constants invariant:
\be\label{TRICOND}
\delta C_{ijk}=\theta_{il}C_{ljk}+\theta_{jl}C_{ilk}+\theta_{kl}C_{ijl}=0~~\Rightarrow~~Q_{ijkl}\theta_{kl}=-2\theta_{ij}.
\ee
Since for $\R,\C$ the structure constants are zero and since for $\Q$ the Levi-Civita symbol is indeed invariant under $\mathfrak{so}(3)$, it is obvious that the algebra of derivations is $\mathfrak{aut}(\Al)\cong\varnothing,\varnothing,\mathfrak{so}(3)$ for $\Al=\R,\C,\Q$ respectively. However, condition $(\ref{TRICOND})$ is not trivially satisfied when $\Al=\Oct$, introducing  seven constraints, one for each quadrangle in $(\ref{TRICOND})$, on the $21$ parameters of $\mathfrak{so}(7)$ leading to the $14$-dimensional algebra $\mathfrak{g}_2$. An equivalent way of thinking about the $\mathfrak{g}_2$ transformation is the following: We are interested in the defining representation $\rep{7}$ of $\mathfrak{g}_2$ so it is most convenient to start with the vector of $\mathfrak{so}(7)$ which is already a $\rep{7}$. This means that one should start with the $\mathfrak{so}(7)$ transformation and project out the constraints on the parameters. This can be implemented by introducing a real parameter $x$ and writing:
\be
\delta V=-\frac{1}{4}\theta_{ij}\Big(-x(e_ie_j)V+xV(e_ie_j)+[e_i,e_j,V]\Big)
\ee
Clearly this expression gives the desired $\delta V_i=-\theta_{ij}V_j$ when $x=1$. Now we demand that we still get $\delta V_i=-\theta_{ij}V_j$ for $x\neq1$ and we arrive at the condition $Q_{ijkl}\theta_{kl}=-2\theta_{ij}$. This is true for any $x$, but it is common to use $x=2/3$ reflecting the fact that $1/3$ of the independent parameters was projected out:
\be
\delta V=-\frac{1}{4}\theta_{ij}\Big(-\frac{2}{3}(e_ie_j)V+\frac{2}{3}V(e_ie_j)+[e_i,e_j,V]\Big)=\frac{2}{3}\vartheta V-\frac{2}{3}V\vartheta+\hat{a}V:=\hat{d}V,
\ee
where the final notation is to reflects the fact that $\mathfrak{g}_2$ corresponds to the algebra of derivations of the octonions. Since all vector, spinor and conjugate spinors of $\mathfrak{so}(8)$ decompose to the $\rep{7}$ of $\mathfrak{g}_2$ we can substitute $\hat{a}$ back into our $\mathfrak{so}(8)$ transformations and express them in a \emph{derivations manifest} form which will be useful when we come to study $\mathfrak{tri}(\Al)$ next:
\begin{align}\label{DERIV}
\delta V&=\alpha V+V\beta+\hat{d}V,&&\alpha:=\frac{1}{2}\theta_{0i}e_i+\frac{1}{12}\theta_{ij}(e_ie_j)&&\\
\delta\psi^*&=(\beta-\alpha)\psi^*+\psi^*\beta+\hat{d}\psi^*,&&\beta:=\frac{1}{2}\theta_{0i}e_i-\frac{1}{12}\theta_{ij}(e_ie_j)&&\\
\delta\chi&=\alpha\chi+\chi(\alpha-\beta)+\hat{d}\chi
\end{align}
This form reflects the fact that the transformation of any of the three eight-dimensional representations  of $\mathfrak{so}(8)$ can be expressed in the form \textit{multiplication from the left} + \textit{multiplication from the right} + \textit{derivation}, with  triality relating them.

\subsubsection{$\mathfrak{tri}(\Al)$}

The triality algebra \cite{Ramond:1976aw} can be thought of as the largest Lie algebra  acting on $\Al$ while preserving the relation $V=\chi\psi^*$. It is defined as:
\be\label{DEFTRI}
\mathfrak{tri}(\Al):=\{(\hat{A},\hat{B},\hat{C})\in3\mathfrak{so}(n)~|~\hat{A}(xy)=x(\hat{B}y)+(\hat{C}x)y, \; x,y\in\Al_n \}.
\ee
A priori one must treat all $(xy,x,y)$ as vectors, perform three independent $\mathfrak{so}(n)$ transformations and identify constraints imposed by the defining property. Comparing with the known transformations studied in the previous sections we find that $(xy,x,y)$ can be identified with $(V,\chi,\psi^*)$. The results are summarised in \autoref{TRITABLE}.
\begin{table}[h!]
\centering
\footnotesize
\begin{tabular}{C{1cm} L{2.5cm} C{2cm} C{2.5cm} C{2.8cm} C{2.5cm}}
\hline\hline
\\
$\Al$ & Conditions & $Tri(\Al)$ & $\delta V$ & $\delta\psi^*$ & $\delta\chi$ 
\\ \\
\hline \\
$\Oct$ & $\hat{d}_A=\hat{d}_B=\hat{d}_C:=\hat{d}$  & $\mathfrak{so}(8)$ & $\alpha V+V\beta+\hat{d}V$  & $(\beta-\alpha)\psi^*+\psi^*\beta+\hat{d}\psi^*$ & $\alpha\chi+\chi(\alpha-\beta)+\hat{d}\chi$ 
\\
& $\alpha_C=\alpha_A:=\alpha$
\\
& $\beta_B=\beta_A:=\beta$
\\
& $\alpha_B=\beta-\alpha$
\\
& $\beta_C=\alpha-\beta$
\\ \\
\hline \\
$\Q$ & $\theta_C=\theta_A:=\theta$ & $3\mathfrak{su}(2)$ & $\theta V+V\theta$ & $\Theta\psi^*+\psi^*\bar\theta$  & $\theta\chi-\chi\Theta$ 
\\
& $\bar\theta_B=\bar\theta_A:=\bar\theta$
\\
& $\bar\theta_C=-\theta_B:=-\Theta$
\\ \\
\hline \\
$\C$ & $\theta_B:=\tfrac{1}{2}(\theta+\theta')$ & $2\mathfrak{u}(1)$ & $2\theta V$ & $(\theta+\theta')\psi^*$  & $(\theta-\theta')\chi$
\\
& $\theta_C:=\tfrac{1}{2}(\theta-\theta')$
\\
& $\theta_A=\theta_B+\theta_C=\theta$
\\ \\
\hline\hline
\end{tabular}
\caption{Vector, spinor and conjugate spinor representations of $\mathfrak{tri}(\Al)$.}\label{TRITABLE}
\end{table}

\subsubsection{$\mathfrak{sa}(N;\Al)$}

The goal of this subsection is to show how $\R,\C,\Q$ allow for a single definition to cover all compact classical Lie algebras and how the natural extensions of the definition to include $\Oct$ will yield the exceptional Lie algebra $\mathfrak{f}_4$. The use of anti-Hermitian generators for the orthogonal and unitary groups is extended naturally to the symplectic ones using the quaternions:
\be
\mathfrak{a}(N;\Al):=\{T\in\Al[N]~|~T^\dagger=-T\}=\begin{cases}
\mathfrak{so}(N) \quad\quad &\text{for }n=1\\
\mathfrak{u}(N) \quad\quad &\text{for }n=2\\
\mathfrak{usp}(2N) \quad\quad &\text{for }n=4
\end{cases}
\ee
One might wonder whether it is possible to find a similar definition for the special unitary algebras, instead of the unitary algebras. Simply using traceless generators does the trick for $\mathfrak{su}(N)$ but  truncates too many generators for $\mathfrak{usp}(2N)$, so we need to compensate accordingly. It turns out that the correct definition indeed uses the traceless anti-hermitian matrices:
\be
\mathfrak{a}'(N;\Al):=\{T\in\Al[N]~|~T^\dagger=-T,\text{Tr}T=0\},
\ee
so that
\be
\text{dim}\ \mathfrak{a}'(N;\Al)=\frac{N(N-1)}{2}+\frac{(n-1)N(N+1)}{2}-(n-1).
\ee
The $(n-1)(n-2)/2=3$ missing generators in quaternionic case are compensated by including the algebra of derivations,
\be
\mathfrak{sa}(N;\Al):=\mathfrak{a}'(N;\Al)+\mathfrak{aut}(\Al)\cong
\mathfrak{a}'(N;\Al)+\mathfrak{so}(n-1)=\begin{cases}
\mathfrak{so}(N) \quad\quad &\text{for }n=1\\
\mathfrak{su}(N) \quad\quad &\text{for }n=2\\
\mathfrak{usp}(2N) \quad\quad &\text{for }n=4
\end{cases}
\ee
 However, it is not obvious which of the two definitions is appropriate in the octonionic case since $\mathfrak{aut}(\Al)=\mathfrak{g}_2\ncong\mathfrak{so}(7)$ and thus the possible compensating terms are inequivalent. Luckily there is a better definition  for $\mathfrak{sa}(N;\Al_n)$ which can unambiguously be extended to include the octonions \cite{Sudbery:1984}.

Firstly, we need to introduce the  \textit{Jordan algebras} \cite{Jordan:1933vh}. A Jordan algebra $\mathcal{J}$ is a commutative but not necessarily associative algebra equipped with a product rule $\ast$ satisfying the Jordan identity:
\be
(X\ast Y)\ast X^2=X\ast(Y\ast X^2),\hspace{2cm}X,Y\in\mathcal{J}.
\ee
The set   $\lieh_N(\Al)$ of $N\times N$ Hermitian  matrices over $\Al$   forms a Jordan algebra under the the product rule
\be
H\ast K:=\frac{1}{2}(HK+KH),\hspace{2cm}H,K\in\lieh_N(\Al),
\ee
given $n=1,2,4$. In the octonionic case, the product definition will still satisfy the Jordan identity for  $N\leq3$. It turns out that the algebra of derivations of these matrices is isomorphic to the classical Lie algebras and can be taken as their definition:
\be
\mathfrak{sa}(N;\Al):= \mathfrak{aut}(\lieh_N(\Al))\cong\begin{cases}
\mathfrak{so}(N)\cong\mathfrak{a}'(N;\R) \quad\quad &\text{for }n=1\text{ and general }N\\
\mathfrak{su}(N)\cong\mathfrak{a}'(N;\C) \quad\quad &\text{for }n=2\text{ and general }N\\
\mathfrak{usp}(2N)\cong\mathfrak{a}'(N;\Q)+\mathfrak{usp}(2) \quad\quad &\text{for }n=4\text{ and general }N\\
\mathfrak{so}(9)\cong\mathfrak{a}'(2;\Oct)+\mathfrak{so}(7) \quad\quad &\text{for }n=8\text{ and }N=2\\
\mathfrak{f}_{4(-52)}\cong\mathfrak{a}'(3;\Oct)+\mathfrak{g}_2 \quad\quad &\text{for }n=8\text{ and }N=3
\end{cases}
\ee
Notice the accidental isomorphism $\mathfrak{a}'(2;\Oct)+\mathfrak{so}(7)\cong\mathfrak{a}'(9;\R)$.

\subsubsection{$\mathfrak{sl}(N;\Al)$}

The above definition can be further extended to the special linear algebra \cite{Sudbery:1984}. The classical Lie algebras $\mathfrak{so}(N),\mathfrak{su}(N),\mathfrak{usp}(2N)$ can be thought as compact subalgebras of the Lie algebras $\mathfrak{sl}(N;\Al)$ for $n=1,2,4$, respectively, such that the non-compact generators are given by $N\times N$ Hermitian traceless matrices over $\Al$ denoted as $\lieh'_N(\Al)$. Now we can use our general definition of $\mathfrak{sa}(N;\Al)$ and define \cite{Barton:2003}:
\be\label{SLAL}
\mathfrak{sl}(N;\Al):=\mathfrak{sa}(N;\Al)+\lieh'_N(\Al)\cong\begin{cases}
\mathfrak{sl}(N;\R) \quad\quad &\text{for }n=1\text{ and general }N\\
\mathfrak{sl}(N;\C) \quad\quad &\text{for }n=2\text{ and general }N\\
\mathfrak{su}^*(2N) \quad\quad &\text{for }n=4\text{ and general }N\\
\mathfrak{so}(1,9) \quad\quad &\text{for }n=8\text{ and }N=2\\
\mathfrak{e}_{6(-26)} \quad\quad &\text{for }n=8\text{ and }N=3
\end{cases}
\ee
In fact, relation $(\ref{SLAL})$ leads to the accidental isomorphism $\mathfrak{sl}(2;\Al_n)\cong\mathfrak{so}(1,n+1)$ explaining why it so natural to study off-shell sYM theories in $D=n+2$.

\subsubsection{$\mathfrak{sp}(2N;\Al)$}

For the sake of completeness we present how the extension of $\lie{a}'(N;\Al)$ to $\mathfrak{sa}(N;R)$ can be applied to extend the standard definition of the symplectic algebra \cite{Barton:2003}. The standard definition covering $\R$ and $\C$ involves the generators:
\be
\lie{sp}'(2N;\Al):=\{T\in\Al[2N]~|~T\Omega+\Omega T^\dagger=0,\text{Tr}T=0\},
\ee
where
\be
\text{dim}\ \lie{sp}'(2N;\Al) =\frac{2N(N+1)}{2}+\frac{2(n-1)N(N-1)}{2}+nN^2-(n-1).
\ee
Adding the missing generators, $\mathfrak{aut}(\Al_n)$, as above we arrive at the desired result for all $\Al$:
\be
\mathfrak{sp}(2N;\Al)\cong\begin{cases}
\mathfrak{sp}(2N;\R)\cong\lie{sp}'(2N;\R) \quad\quad &\text{for }n=1\text{ and general }N\\
\mathfrak{su}(N,N)\cong\lie{sp}'(2N;\C) \quad\quad &\text{for }n=2\text{ and general }N\\
\mathfrak{so}^*(4N)\cong\lie{sp}'(2N;\Q)+\mathfrak{usp}(2) \quad\quad &\text{for }n=4\text{ and general }N\\
\mathfrak{so}(2,10)\cong\lie{sp}'(2N;\Oct)+\mathfrak{so}(7) \quad\quad &\text{for }n=8\text{ and }N=2\\
\mathfrak{e}_{7(-25)}\cong\lie{sp}'(2N;\Oct)+\mathfrak{g}_2 \quad\quad &\text{for }n=8\text{ and }N=3
\end{cases}
\ee

\subsubsection{The Magic Square}

The magic square \cite{Freudenthal:1954,Tits:1955, Rosenfeld:1956, Freudenthal:1959, Tits:1966}  is a  well-studied construction appearing in  various mathematical \cite{Barton:2003, Landsberg200259, westbury2006sextonions, landsberg2006sextonions} and physical contexts \cite{Julia:1980gr, Gunaydin:1983rk, Gunaydin:1983bi, Gunaydin:1984ak, Gunaydin:1992zh, Bellucci:2006xz, Gunaydin:2007bg, Borsten:2008wd, Gunaydin:2009zza, Marrani:2012uu, Borsten:2012fx, Cacciatori:2012cb, Chiodaroli:2015wal, Borsten:2017uoi} and the references therein. Although we will not build it here, we will sketch how the commutation relations can be studied by looking at the particular example associated with $\mathfrak{f}_4$. The defining representation, $\rep{26}$, can be represented as an element in $\lieh'_3(\Oct)$:
\be
H=\begin{pmatrix}
h_1 & \psi^* & \chi^* \\
\psi & h_2 & V^* \\
\chi & V & h_3
\end{pmatrix},~~~V,\psi,\chi,\in\Oct,~~~h_1,h_2,h_3=-h_1-h_2\in\R.
\ee
Since $\mathfrak{f}_4\cong\mathfrak{a}'(3;\Oct)+\mathfrak{g}_2$ the action on $H$ is given by,
\be\label{F4TRA}
\delta H=[T,H]+\hat{d}H
\ee
where $T$ is an element of $\mathfrak{a}'(3;\Oct)$ and can be written as \be
T=\begin{pmatrix}
\beta-\alpha & -x_s^* & x_c^* \\
x_s & -\beta & -x_v^* \\
-x_c & x_v & \alpha
\end{pmatrix},~~~x_v,x_s,x_c\in\Oct,~~~\alpha,\beta\in\text{Im}\Oct.
\ee
Splitting $T$ into its off-diagonal part $X$ and the diagonal part $A$ and substituting back in equation $(\ref{F4TRA})$ the transformation becomes:
\be\label{MAGICTRA}
\delta H=\Big([X,H]\Big)+\Big([A,H]+\hat{d}H\Big)=:[X,H]+\hat{M}H.
\ee
By turning off $X$ for the moment and calculating the explicit variation of each entry in $H$ one finds that the second part in $(\ref{MAGICTRA})$ corresponds exactly to an $\mathfrak{so}(8)$ transformation according to  \autoref{TRITABLE} and, hence,  $\mathfrak{f}_4\cong3\Oct+\mathfrak{tri}(\Oct)$.  Alternatively, we could have started with the algebra $\mathfrak{tri}(\Oct)+3\Oct$ and, by repeated variations, deduced all commutation relations, which would  indeed confirm it is  isomorphic to $\mathfrak{f}_4$. The above process can be repeated for any of the division algebras leading to:
\be
\mathfrak{sa}(3;\Al)\cong\mathfrak{tri}(\Al)+3\Al.
\ee
One can follow the exact same procedure but now allowing for $\hat{M}\otimes\hat{1}+\hat{1}\otimes\hat{\tilde{M}}$ and $X\otimes\tilde{X}$ leading to the magic square formula in Barton-Sudbery form \cite{Barton:2003}:
\be
\mathfrak{M}(\Al,\tilde\Al)=\mathfrak{tri}(\Al)\oplus\mathfrak{tri}(\tilde\Al)+3(\Al\otimes\tilde\Al).
\ee
The commutation relations between the $X\otimes\tilde{X}$ generators come with a free parameter $w=\pm1$ which distinguishes between the compact and a particular non-compact real form of the complexified algebra \cite{Anastasiou:2013hba}. In this review we are interested in the latter choice which leads to the Lorentzian FRT magic square as in \autoref{LFRT}.

\section{The Magic Pyramid of Supergravities}

In this last section we will first study how the relations in the previous section lead to a very natural formulation of sYM theories in $D=n+2$ over the NDAs \cite{Kugo:1982bn, Baez:2009xt, Anastasiou:2013cya}\footnote{Note, supergravity in $D=n+3$  admits a similar division-algebraic formulation \cite{Baez:2010ye, Anastasiou:2014zfa}.}. We will then use this formulation to assign a physical interpretation to the magic square formula, which will allow us to extend it to all $D=n+2$, thus building the so-called magic pyramid of supergravities. 

\subsection{Super Yang-Mills in $D=n+2$ over the Division Algebras}

The theories we wish to study are listed in \autoref{11}.
We start this section with a few observations to motivate the formulation of this class of sYM theories over the NDAs:
\begin{itemize}
\item It is obvious from \autoref{11} that the NDAs have a dual role. The first is related to the spacetime dimension $D=n+2$ and to the fact that in these dimensions the on-shell, i.e.~$\mathfrak{so}(n)$, vector and spinor representation spaces are given by $\rep{v}(n)\cong\rep{s}(n)\cong\Al_n$. The second role is related to the number of supercharges $\charge=2n'$ and to the fact that the total bosonic and fermionic spaces are given by $\rep{v}(n)\oplus\mathds{R}^{n'-n}\cong\rep{s}(n)^\susy\cong\Al'$. Simply put $\Al'$ is the algebra in which the theory is formulated, but only once  the spacetime subalgebra $\Al$ is fixed  can we interpret the degrees of freedom correctly using the  relation $n'=\susy~n$ \cite{ Anastasiou:2013cya}. Note, for spacetime dimensions $D=n+2$ the minimal $\mathfrak{so}(1, n+1)$ spinor representations are given by  $\mathds{A}_n$-doublets, so that $\mathcal{Q}=\mathcal{N}\times 2n$, in agreement with  \autoref{2}.  
\item The minimal theories in $D=n+2$ labelled as \cir{1}, \cir{3}, \cir{6}, \cir{7} have a very natural description over the NDAs \cite{Kugo:1982bn, Baez:2009xt, Anastasiou:2013cya}. The first hint comes from the isomorphism $\mathfrak{sl}(2;\Al)\cong\mathfrak{so}(1,n+1)$ \cite{Sudbery:1984}, which implies that the usual off-shell formulation of the minimal theory in $D=4$ for the complexes can be generalised to all $\Al$. Furthermore, these are the only theories where $\Al$ and $\Al'$ are the same. This is due to the fact that there is a single spinor and no scalars and so the interpretation between spacetime and internal degrees of freedom is redundant. Finally we observe that the full global symmetry algebra in these cases is $\mathfrak{tri}(\Al)$. 
\item The $D=3$ theories have been assigned labels identical to the ones of which they are descendants. This is because they only differ on interpretation: in the higher-dimensional theory the degrees of freedom are all spacetime while in the $D=3$ descendants they are all internal. Simply put, the   theories with $\Al\cong\Al'$ have a single vector and a single spinor valued in $\Al'$ while their $D=3$ descendants have $n'$ scalars and $n'$ spinors, both valued in $\R$.  
\end{itemize} 
With these observations it is obvious that as far as the symmetries are concerned all the hard-work has been done in the previous section. Since we have the simple transformation rules for \cir{1}, \cir{3}, \cir{6}, \cir{7} all we have to do is provide a consistent interpretation for the remaining theories and understand how their transformations follow. The resulting picture is  described in \autoref{12}, the content of which can be summarised by the following expressions:
\begin{align}
&\rep{v}(n)\cong\Al_{n-\delta_{n1}}\\
&\rep{S}(n,n')\cong\R^{n'-n+\delta_{n1}}\\
&\rep{s}(n)\cong\Al\\
&\rep{v}(n)\oplus\rep{S}(n,n')\cong\rep{s}(n)^{\susy}\cong\Al_{n'}\\
&\susy=n'/n\\
\nonumber\\
&\lie{st}(n)=\lie{so}(n)\\
&\lie{int}(n,n')=\lie{a}(\susy;\mathds{D})\oplus\lie{q}(n,n')\\
&\lie{q}(n,n')=\delta_{n1}\delta_{n'4}\lie{so}(3)\oplus\delta_{n1}\delta_{n'2}\lie{so}(2)\ominus\delta_{n2}\delta_{n'8}\lie{u}(1)
\end{align}
Here, we have used $\lie{st}$ to denote the spacetime little group algebra. The Kronecker deltas reflect the fact that in $D=3$ the vector dualises to scalar leading to an enhancement of the $R$-symmetry and that in $D=4$ the maximal multiplet is self-conjugate and thus cannot support the $\mathfrak{u}(1)$ of the $R$-symmetry.

\begin{table}[h!]
\centering
\footnotesize
\begin{tabular}{C{1cm} | C{0.25cm} L{3cm} L{3.5cm} L {3cm} L{3cm}}
\hline\hline
\\
$D$~~ && $\charge=16$ & $\charge=8$ & $\charge=4$ & $\charge=2$
\\ \\
\hline \\
$10$ && \cir{1} : $(\Oct,\Oct')$ : $\mathfrak{so}(8)$
\\ \\
$6$ && \cir{2} : $(\Q,\Oct')$ : $4\mathfrak{usp}(2)$ & \cir{3} : $(\Q,\Q')$ : $3\mathfrak{usp}(2)$
\\ \\
$4$ && \cir{4} : $(\C,\Oct')$ : $\mathfrak{u}(4)$ & \cir{5} : $(\C,\Q')$ : $\mathfrak{u}(2)\oplus \mathfrak{u}(1)$ & \cir{6} : $(\C,\C')$ : $2\mathfrak{u}(1)$
\\ \\
$3$ && \cir{1} : $(\R,\Oct')$ : $\mathfrak{so}(8)$ & \cir{3} : $(\R,\Q')$ $3\mathfrak{so}(3)$ & \cir{6} : $(\R,\C')$, $2\mathfrak{so}(2)$ & \cir{7} : $(\R,\R')$ : $\emptyset$
\\ \\
\hline\hline
\end{tabular}
\caption{The super-Yang-Mills theories in $D=n+2$. They are labelled with circled numbers for ease of reference. The labelling in terms of division algebras is explained below. We have included the global symmetries of the on-shell theories as a direct sum of the global on-shell spacetime algebra $\mathfrak{so}(D-2)$ and the global internal algebra $\mathfrak{int}$.}\label{11}
\end{table}
\begin{table}[h!]
\footnotesize
\hspace{-2cm}\begin{tabular}{C{0.2cm} | C{0cm} L{4cm} C{0cm} L{4cm} C{0cm} L{4cm} C{0cm} L{4cm}}
\hline\hline
\\
$D$~~ && $\charge=16$ && $\charge=8$ && $\charge=4$ && $\charge=2$
\\ \\
\hline \\
$10$ && $\cir{1} : (\Oct,\Oct')$ \\
&& $\rep{v}(8)\cong\Oct$, $\rep{s}(8)\cong\Oct$ \\
&& $\lie{st}(8)=\lie{so}(8)$ \\
&& $V : \rep{8}_v$, $\psi : \rep{8}_s$ \\ \\
&& The theory is formulated in $\Oct'$ so full Fano plane. Both the vector and the spinor are full octonions. \\
\\ \\
$6$ && $\cir{2} : (\Q,\Oct')$ && $\cir{3} : (\Q,\Q')$ \\ 
&& $\rep{v}(4)\cong\Q$, $\rep{S}(4,8)\cong\Q$, && $\rep{v}(4)\cong\Q$, \\ 
&& $\rep{s}(4)\cong\Q$ && $\rep{s}(4)\cong\Q$ \\
&& $\lie{st}(4)=2\lie{sp}(1)$, && $\lie{st}(4)=2\lie{sp}(1)$, \\
&& $\lie{int}(4,8)=2\lie{sp}(1)$ && $\lie{int}(4,4)=\lie{sp}(1)$ \\
&& $V : (\rep{2,2;1,1})$, $\phi : (\rep{1,1;2,2})$, && $V : (\rep{2,2;1})$, \\
&& $\psi : (\rep{2,1;2,1})+(\rep{1,2;1,2})$ && $\psi : (\rep{2,1;2})$ \\ \\
&& The theory is formulated in && The theory is formulated in \\ 
&& $\Oct'$ so again full Fano plane. && $\Q'$ and it can be obtained in \\
&& Both the vector and a single && two ways. The first is to start \\
&& spinor are quaternions so && with \cir{1} and simply use \\
&& a quaternionic subspace of the && quaternions instead of \\
&& plane is playing the role of && octonions. The second is to \\
&& spacetime while the remaining && start with \cir{2} and simply \\
&& bases are internal. Therefore && throw away the internal \\ 
&& the scalars are an internal && quadrangle of the \\
&& quaternion while all the && Fano plane.\\
&& fermionic dof form a complex \\
&& pair of quaternionic spinors.   \\ \\
$4$ && $\cir{4} : (\C,\Oct')$ && $\cir{5} : (\C,\Q')$ && $\cir{6} : (\C,\C')$ \\ 
&& $\rep{v}(2)\cong\C$, $\rep{S}(2,8)\cong\R^6$, && $\rep{v}(2)\cong\C$, $\rep{S}(2,4)\cong\C$, && $\rep{v}(2)\cong\C$,\\ 
&& $\rep{s}(2)\cong\C$ && $\rep{s}(2)\cong\C$ && $\rep{s}(2)\cong\C$\\
&& $\lie{st}(2)=\lie{u}(1)$, $\lie{int}(2,8)=\lie{su}(4)$  && $\lie{st}(2)=\lie{u}(1)$, $\lie{int}(2,4)=\lie{u}(2)$ && $\lie{st}(2)=\lie{u}(1)$, $\lie{int}(2,2)=\lie{u}(1)$ \\
&& $V : \rep{1}_{-2}+\rep{1}_2$, $\phi : \rep{6}_0$, && $V : \rep{1}_{-2}^0+\rep{1}_2^0$, $\phi : \rep{1}_{0}^{2}+\rep{1}_0^{-2}$, && $V : (-2,0)+(2,0)$\\
&& $\psi : \rep{4}_{-1}+\brep{4}_1 $ && $\psi : \rep{2}_{-1}^{1}+\rep{2}_{1}^{-1}$ && $\psi : (-1,1)+(1,-1)$ \\ \\
&& The theory is formulated in &&  This is most easily obtained as && This theory is a truncation of \\
&& $\Oct'$ so again a full Fano plane. && a truncation of \cir{4}. Throwing && \cir{4} by throwing away all six \\
&& Both the vector and a single && away a quadrangle leaves only && imaginary internal bases, with \\
&& spinor are complex numbers && two internal imaginary bases. && effect of truncating out all \\
&& so we choose one imaginary && Therefore we are left with a && scalars and leaving only a \\
&& base of the Fano plane as && complex scalar while the && single complex spinor. This \\
&& spacetime. The remaining six && fermionic dof are a complex && theory can also be obtained \\
&& bases are internal and thus we && pair of complex spinors. && by starting from \cir{1} and using \\
&& have six real scalars. The &&&& complexes instead of \\
&& fermionic dof form a &&&& octonions. \\
&& quaternionic pair of complex &&&& \\
&& spinors. \\ \\
$3$ && $\cir{1}: (\R,\Oct')$ && $\cir{3} : (\R,\Q')$ && $\cir{6} : (\R,\C')$ && $\cir{7} : (\R,\R')$ \\ 
&& $\rep{S}(1,8)\cong\Oct$, $\rep{s}(1)\cong\R$ && $\rep{S}(1,4)\cong\Q$, $\rep{s}(1)\cong\R$ && $\rep{S}(1,2)\cong\C$, $\rep{s}(1)\cong\R$ && $\rep{S}(1,1)\cong\R $\\ 
&& $\lie{int}(1,8)=\lie{so}(8)$  && $\lie{int}(1,4)=3\lie{so}(3)$ && $\lie{int}(1,2)=2\lie{so}(2)$&& $\rep{s}(1)\cong\R $\\  
&& $\phi : \rep{8}_v$, $\psi : \rep{8}_s$, && $\phi : \rep{(2,2,1)}$, $\psi : \rep{(1,2,2)}$, && $\phi : (-2,0)+(2,0)$ && $\lie{int}(1,2)=\emptyset$ \\ 
&&&&&& $\psi : (-1,1)+(1,-1)$ && $\phi : 1$, $\psi : 1$ \\ \\
\hline\hline
\end{tabular}
\caption{sYM in $D=n+2$ and their Fano plane interpretation.}\label{12}
\end{table}

\FloatBarrier

\subsection{Magic pyramid formula}

The results of the previous subsection allow us to give a more physical version of the magic square formula, one that explains why the Lie algebras correspond to the the U-duality algebras of supergravity theories in $D=3$:
\be
\begin{split}
\lieg(1,n',\tilde{n}')=&\Big[\mathfrak{int}(1,n')\oplus\mathfrak{int}(1,\tilde{n}')+\rep{s}(1)^\susy\otimes\rep{s}(1)^{\tilde{\susy}}\Big]\\&+\Big[\rep{s}(1)^\susy\otimes\rep{s}(1)^{\tilde{\susy}}+\rep{S}(1,n')\otimes\rep{S}(1,\tilde{n}')\Big]
\end{split}
\ee
where the first and second square brackets correspond to the maximal compact subalgebra $\lie{h}$ and the non-compact complement $\lie{p}$, respectively. The formula corresponds to the decomposition of the adjoint representation of $\mathfrak{g}$ under $\mathfrak{g}\supset \mathfrak{h}\supset \mathfrak{int}\oplus \tilde{\mathfrak{int}}$. The fact that this is the unique decomposition implies that the knowledge of the decomposed form together with the compact and non-compact parts is enough to deduce $\mathfrak{g}$.

In order to extend the magic square of supergravities to a magic pyramid of supergravities  we extend the formula  to all $n$. The correct extension is determined by the non-compact generators: as mentioned earlier the scalars of the theory parametrise the symmetric coset $\mathcal{G}/\mathcal{H}$, therefore understanding the squaring origin of the scalars will give  the extension. In general
there are three terms from which scalars can result and we will now study
each one of them separately:

\paragraph{$V\circ \tilde{V}$} This is the on-shell tensor product between the
\textit{Left} and \textit{Right} vector fields which decomposes into a symmetric
traceless part, an antisymmetric part and a trace part with the scalar d.o.f.~corresponding to the latter. Since both vector fields are singlets under
their respective global internal symmetries, the resulting scalar will be
a singlet as well; therefore, since our formulae denote representation spaces with
respect to $\mathfrak{int}(n,n')\oplus \mathfrak{int}(n,\tilde{n}')\times\delta_{n2}\mathfrak{u}(1)_d$, the scalar from this
tensor product will simply contribute as $\R\otimes\R$. However, in $D=3$,
we have dualised the vector into a scalar and thus this dof has already been
taken into account in the scalar vector space; in order to avoid double counting this term will come with a coefficient
$(1-\delta_{n1})$. Furthermore, we should note that in $D=4$ the on-shell
$2$-form potential corresponding to the antisymmetric part of the tensor
product dualises to a scalar and thus this contributes an extra $\delta_{n2}\R\otimes\R$
term. Putting everything together, the vector $\otimes$ vector contribution
to the scalar coset space is:
\be
(1-\delta_{n1})\R\otimes\R+i\delta_{n2}\R\otimes\R.
\ee
The $i$ next to the second term is to highlight the fact that in $D=4$ the
two terms will always carry opposite charges with respect to $\mathfrak{u}(1)_d$.
\paragraph{$\psi\circ\tilde\psi$} The second scalar contribution comes from the
on-shell tensor product between the spinors of each sYM theory. Upon tensoring
two spinors the tensor space $\rep{s}(n)\otimes\rep{s}(n)$ decomposes
on the Clifford-algebra basis with each coefficient corresponding to a $p$-form
potential; as a result, the $0$-rank form will contribute to a scalar.
However, some minimal theories may contain spinors of opposite $\mathfrak{u}(1)$ charge or symplectic pairs of spinors, therefore in these cases the product contributes more than one scalar. Thankfully, we have already encoded this in our division-algebra
parametrisation of minimal spinors and therefore the scalar contribution
from the tensor of two minimal spinors is $\mathds{D}[1,1]$. Moreover, since
we are not working only with minimal theories and each of the sYM might have
more than one spinor, the total contribution comes from tensoring
the total spinor spaces $\rep{s}(n)^{\susy}\otimes\rep{s}(n)^{\tilde\susy}$
and therefore the spinor $\otimes$ spinor contribution to the scalar coset
space is:
\be
\mathds{D}[\susy,\tilde\susy],
\ee
which can be thought of as an ${\susy}\times{\tilde\susy}$ matrix
with entries valued over $\mathds{D}$.
\paragraph{$\phi\circ\tilde\phi$} The third and final contribution to the scalar
coset space comes from tensoring the scalars between the \textit{Left} and
\textit{Right} sYM theories. This is the simplest term since the full tensor
product between the total scalar spaces will contribute to scalars and therefore
the last term is simply:
\be
\rep{S}(n,n')\otimes\rep{S}(n,\tilde{n}').
\ee

\begin{figure}[h!]
  \centering
    \includegraphics[width=\textwidth]{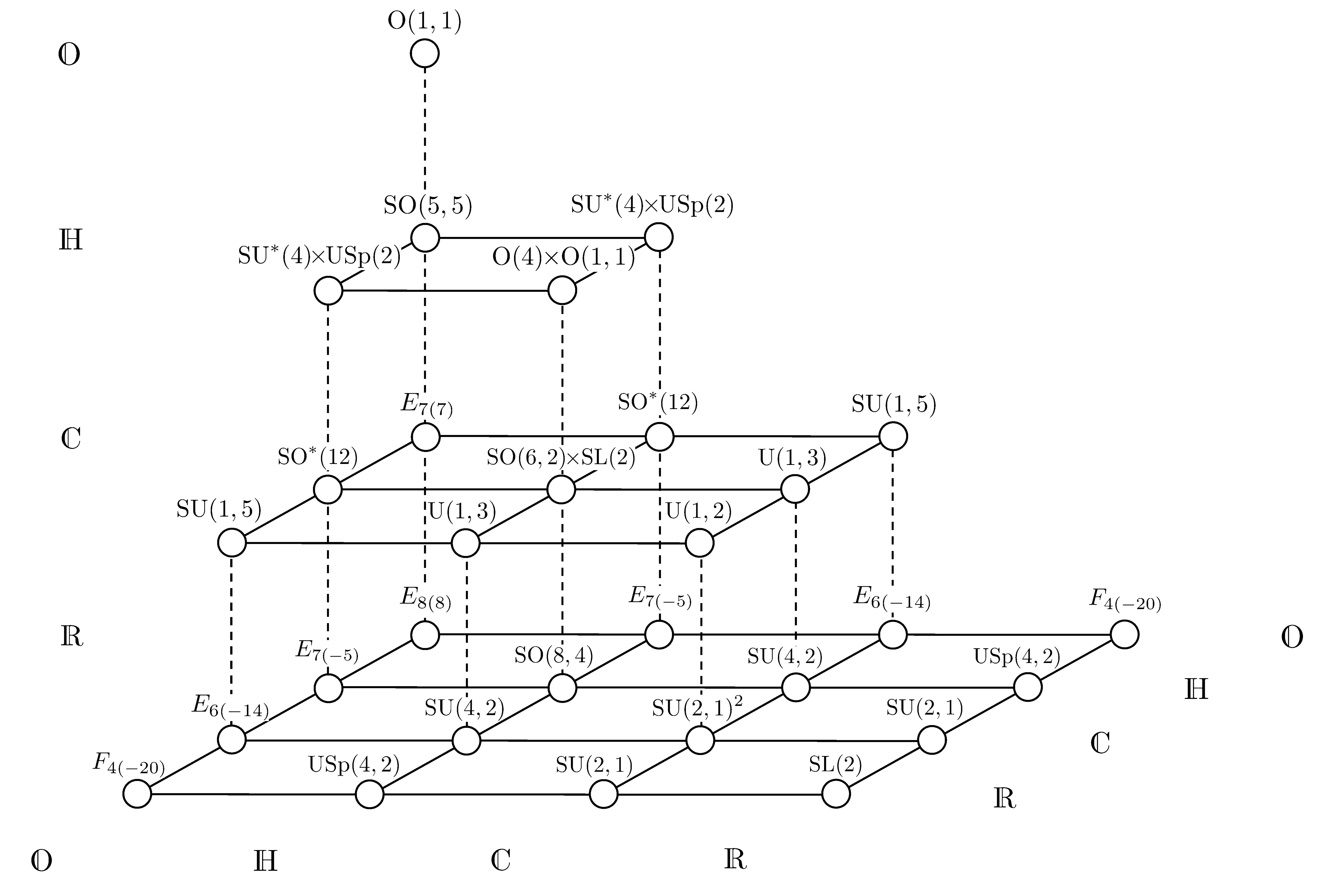}
  \caption[The Fano plane.]{The magic pyramid of U-duality groups obtained from the product of two sYM theories in $D=3,4,6,10$ \cite{Anastasiou:2013hba}.}\label{mp}
\end{figure}

Putting all this together we arrive at the magic pyramid formula:
\be
\begin{split}
\lieg(n,n',\tilde{n}')=&\Big[\mathfrak{int}(n,n')\oplus\mathfrak{int}(n,\tilde{n}')\oplus\delta_{n2}\lie{u}(1)_d+\mathds{D}[\susy,\tilde\susy]\Big]\\&+\Big[(1-\delta_{n1})\R\otimes\R+i\delta_{n2}\R\otimes\R+\mathds{D}[\susy,\tilde\susy]+\rep{S}(n,n')\otimes\rep{S}(n,\tilde{n}')\Big],
\end{split}
\ee
where the complete set of commutation relations are given in \cite{Anastasiou:2015vba}. One can now apply the formula for all possible values of $(n,n',\tilde{n}')$ and arrive at the magic pyramid of supergravities as in \autoref{mp}.

\section{Conclusion}

We have seen that squaring super Yang-Mills theories in $D=n+2$ dimensions leads to a magic pyramid of supergravity theories, with global symmetries given by the magic pyramid formula, the base of which is the familiar magic square. Note, however, that the vertical axis of the magic pyramid is on a different footing to the horizontal axes in the sense that for $D=6,10$ the exceptional groups do not appear.

\begin{figure}[h!]
  \centering
    \includegraphics[width=\textwidth]{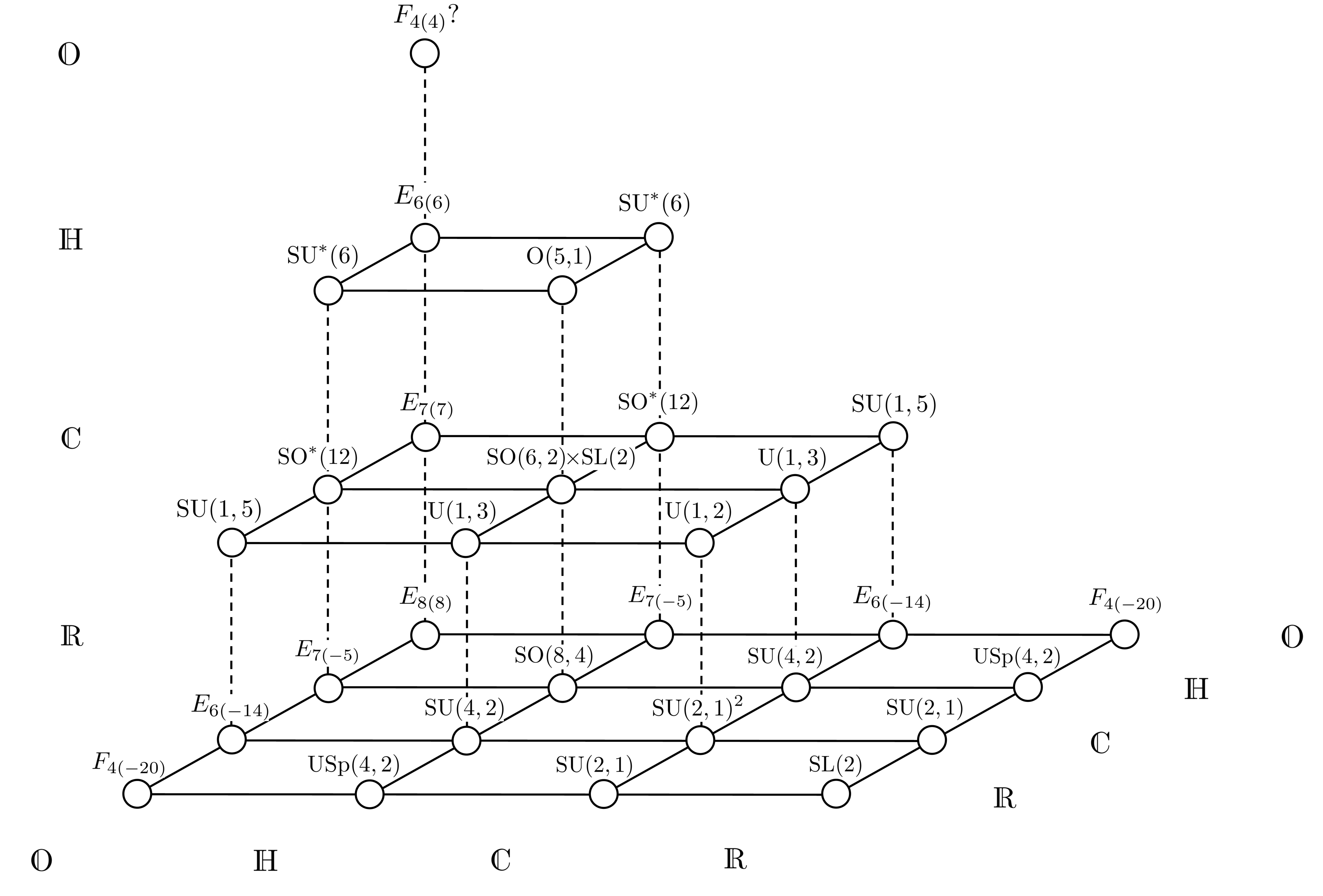}
  \caption[The Fano plane.]{The conformal magic pyramid of U-duality groups obtained from the product of two conformal theories in $D=3,4,6$  \cite{Anastasiou:2013hba}.}\label{CGpyr2}
\end{figure}

Remarkably, this can be remedied in $D=6$ by considering the square of tensor, rather than vector, multiplets. This   gives rise to the so-called \emph{conformal} magic pyramid, given in \autoref{CGpyr2} \cite{Anastasiou:2013hba}. Using the same principles for tracking the squaring origin of scalars we find that, when squaring the maximal tensor multiplets of the same chirality, we get $\mathcal{G}=E_{6(6)}$ corresponding to a conjectured $\susy=(4,0)$ chiral conformal gravitational  theory with a gravi-gerbe field  in place  of the graviton \cite{Hull:2000zn, Hull:2000rr, Chiodaroli:2011pp, Anastasiou:2013hba, Henneaux:2016opm, Borsten:2017jpt, Henneaux:2017xsb}. Simply following the maximal pattern $D=3,4,6$ with $E_{8(8)},E_{7(7)},E_{6(6)}$ suggests that there may be an exotic theory with $\mathcal{G}=F_{4(4)}$ in $D=10$, as suggested in \autoref{CGpyr2}. What this theory might be, if it exists at all, remains to be seen.

%\bibliography{Ref_Library}

\newcommand{\etalchar}[1]{$^{#1}$}
\begin{thebibliography}{BDD{\etalchar{+}}13b}

\bibitem[ABD{\etalchar{+}}14a]{Anastasiou:2013hba}
A.~Anastasiou, L.~Borsten, M.~J. Duff, L.~J. Hughes and S.~Nagy, \textsl{ {A
  magic pyramid of supergravities}},
\newblock JHEP \textbf{ 1404}, 178 (2014), {1312.6523}.

\bibitem[ABD{\etalchar{+}}14b]{Anastasiou:2014zfa}
A.~Anastasiou, L.~Borsten, M.~J. Duff, L.~J. Hughes and S.~Nagy, \textsl{ {An
  octonionic formulation of the M-theory algebra}},
\newblock JHEP \textbf{ 1411}, 022 (2014), {1402.4649}.

\bibitem[ABD{\etalchar{+}}14c]{Anastasiou:2013cya}
A.~Anastasiou, L.~Borsten, M.~J. Duff, L.~J. Hughes and S.~Nagy, \textsl{
  {Super Yang-Mills, division algebras and triality}},
\newblock JHEP \textbf{ 1408}, 080 (2014), {1309.0546}.

\bibitem[ABD{\etalchar{+}}14d]{Anastasiou:2014qba}
A.~Anastasiou, L.~Borsten, M.~J. Duff, L.~J. Hughes and S.~Nagy, \textsl{
  {Yang-Mills origin of gravitational symmetries}},
\newblock Phys.Rev.Lett. \textbf{ 113}(23), 231606 (2014), {1408.4434}.

\bibitem[ABD{\etalchar{+}}17a]{Anastasiou:2016csv}
A.~Anastasiou, L.~Borsten, M.~J. Duff, M.~J. Hughes, A.~Marrani, S.~Nagy and
  M.~Zoccali, \textsl{ {Twin supergravities from Yang-Mills theory squared}},
\newblock Phys. Rev. \textbf{ D96}(2), 026013 (2017), {1610.07192}.

\bibitem[ABD{\etalchar{+}}17b]{Anastasiou:2017nsz}
A.~Anastasiou, L.~Borsten, M.~J. Duff, A.~Marrani, S.~Nagy and M.~Zoccali,
  \textsl{ {Are all supergravity theories Yang-Mills squared?}},
\newblock (2017), {1707.03234}.

\bibitem[ABHN16]{Anastasiou:2015vba}
A.~Anastasiou, L.~Borsten, L.~J. Hughes and S.~Nagy, \textsl{ {Global
  symmetries of Yang-Mills squared in various dimensions}},
\newblock JHEP \textbf{ 148}, 1601 (2016), {1502.05359}.

\bibitem[AHCK10]{ArkaniHamed:2008gz}
N.~Arkani-Hamed, F.~Cachazo and J.~Kaplan, \textsl{ {What is the Simplest
  Quantum Field Theory?}},
\newblock JHEP \textbf{ 09}, 016 (2010), {0808.1446}.

\bibitem[{Alb}06]{Elduque:2006}
{Alberto Elduque}, \textsl{ {A new look at Freudenthal's magic square}},
\newblock Non Associative Algebra and Its Applications, (L. Sabinin, L.
  Sbitneva, I.P. Shestakov, eds.), Lect. Notes in Pure and Applied Mathematics
  \textbf{ 246}, 149--165 (2006).

\bibitem[Bae02]{Baez:2001dm}
J.~C. Baez, \textsl{ {The Octonions}},
\newblock Bull. Am. Math. Soc. \textbf{ 39}, 145--205 (2002), {math/0105155}.

\bibitem[BALW17]{Bahjat-Abbas:2017htu}
N.~Bahjat-Abbas, A.~Luna and C.~D. White, \textsl{ {The Kerr-Schild double copy
  in curved spacetime}},
\newblock (2017), {1710.01953}.

\bibitem[BCD{\etalchar{+}}09]{Bern:2009kd}
Z.~Bern, J.~J. Carrasco, L.~J. Dixon, H.~Johansson and R.~Roiban, \textsl{ {The
  Ultraviolet Behavior of N=8 Supergravity at Four Loops}},
\newblock Phys. Rev. Lett. \textbf{ 103}, 081301 (2009), {0905.2326}.

\bibitem[BCJ08]{Bern:2008qj}
Z.~Bern, J.~Carrasco and H.~Johansson, \textsl{ {New Relations for Gauge-Theory
  Amplitudes}},
\newblock Phys.Rev. \textbf{ D78}, 085011 (2008), {0805.3993}.

\bibitem[BCJ10]{Bern:2010ue}
Z.~Bern, J.~J.~M. Carrasco and H.~Johansson, \textsl{ {Perturbative Quantum
  Gravity as a Double Copy of Gauge Theory}},
\newblock Phys.Rev.Lett. \textbf{ 105}, 061602 (2010), {1004.0476}.

\bibitem[BD15]{Borsten:2015pla}
L.~Borsten and M.~J. Duff, \textsl{ {Gravity as the square of Yang--Mills?}},
\newblock Phys. Scripta \textbf{ 90}, 108012 (2015), {1602.08267}.

\bibitem[BDD{\etalchar{+}}09]{Borsten:2008wd}
L.~Borsten, D.~Dahanayake, M.~J. Duff, H.~Ebrahim and W.~Rubens, \textsl{
  {Black Holes, Qubits and Octonions}},
\newblock Phys. Rep. \textbf{ 471}(3--4), 113--219 (2009), {0809.4685}.

\bibitem[BDD13a]{Bern:2013qca}
Z.~Bern, S.~Davies and T.~Dennen, \textsl{ {The Ultraviolet Structure of
  Half-Maximal Supergravity with Matter Multiplets at Two and Three Loops}},
\newblock Phys. Rev. \textbf{ D88}, 065007 (2013), {1305.4876}.

\bibitem[BDD{\etalchar{+}}13b]{Bern:2013uka}
Z.~Bern, S.~Davies, T.~Dennen, A.~V. Smirnov and V.~A. Smirnov, \textsl{
  {Ultraviolet Properties of N=4 Supergravity at Four Loops}},
\newblock Phys. Rev. Lett. \textbf{ 111}(23), 231302 (2013), {1309.2498}.

\bibitem[BDD14a]{Bern:2014sna}
Z.~Bern, S.~Davies and T.~Dennen, \textsl{ {Enhanced ultraviolet cancellations
  in $\mathcal N=5$ supergravity at four loops}},
\newblock Phys.Rev. \textbf{ D90}(10), 105011 (2014), {1409.3089}.

\bibitem[BDD14b]{Bern:2014lha}
Z.~Bern, S.~Davies and T.~Dennen, \textsl{ {The Ultraviolet Critical Dimension
  of Half-Maximal Supergravity at Three Loops}},
\newblock (2014), {1412.2441}.

\bibitem[BDD{\etalchar{+}}15]{Bern:2013yya}
Z.~Bern, S.~Davies, T.~Dennen, Y.-t. Huang and J.~Nohle, \textsl{
  {Color-Kinematics Duality for Pure Yang-Mills and Gravity at One and Two
  Loops}},
\newblock Phys. Rev. \textbf{ D92}(4), 045041 (2015), {1303.6605}.

\bibitem[BDDH12]{Bern:2012cd}
Z.~Bern, S.~Davies, T.~Dennen and Y.-t. Huang, \textsl{ {Absence of Three-Loop
  Four-Point Divergences in N=4 Supergravity}},
\newblock Phys. Rev. Lett. \textbf{ 108}, 201301 (2012), {1202.3423}.

\bibitem[BDHK10]{Bern:2010yg}
Z.~Bern, T.~Dennen, Y.-t. Huang and M.~Kiermaier, \textsl{ {Gravity as the
  Square of Gauge Theory}},
\newblock Phys.Rev. \textbf{ D82}, 065003 (2010), {1004.0693}.

\bibitem[BDHN14]{Borsten:2013bp}
L.~Borsten, M.~J. Duff, L.~J. Hughes and S.~Nagy, \textsl{ {A magic square from
  Yang-Mills squared}},
\newblock Phys.Rev.Lett. \textbf{ 112}, 131601 (2014), {1301.4176}.

\bibitem[BDL12]{Borsten:2012fx}
L.~Borsten, M.~J. Duff and P.~Levay, \textsl{ {The black-hole/qubit
  correspondence: an up-to-date review}},
\newblock Class.Quant.Grav. \textbf{ 29}, 224008 (2012), {1206.3166}.

\bibitem[BFGM06]{Bellucci:2006xz}
S.~Bellucci, S.~Ferrara, M.~G{\"u}naydin and A.~Marrani, \textsl{ {Charge
  orbits of symmetric special geometries and attractors}},
\newblock Int. J. Mod. Phys. \textbf{ A21}, 5043--5098 (2006),
  {hep-th/0606209}.

\bibitem[BH09]{Baez:2009xt}
J.~C. Baez and J.~Huerta,
\newblock Division Algebras and Supersymmetry I,
\newblock in \textsl{ Superstrings, Geometry, Topology, and C*-Algebras, eds.
  R. Doran, G. Friedman and J. Rosenberg, Proc. Symp. Pure Math}, volume~81,
  pages 65--80, 2009.

\bibitem[BH11]{Baez:2010ye}
J.~C. Baez and J.~Huerta, \textsl{ {Division Algebras and Supersymmetry II}},
\newblock Adv. Theor. Math. Phys. \textbf{ 15}(5), 1373--1410 (2011),
  {1003.3436}.

\bibitem[BHM12]{Bargheer:2012gv}
T.~Bargheer, S.~He and T.~McLoughlin, \textsl{ {New Relations for
  Three-Dimensional Supersymmetric Scattering Amplitudes}},
\newblock Phys.Rev.Lett. \textbf{ 108}, 231601 (2012), {1203.0562}.

\bibitem[BM17]{Borsten:2017uoi}
L.~Borsten and A.~Marrani, \textsl{ {A Kind of Magic}},
\newblock Class. Quant. Grav. \textbf{ 34}(23), 235014 (2017), {1707.02072}.

\bibitem[Bor17]{Borsten:2017jpt}
L.~Borsten, \textsl{ {On $D=6$, $\mathcal{N}=(2,0)$ and $\mathcal{N}=(4,0)$
  theories}},
\newblock (2017), {1708.02573}.

\bibitem[BS03]{Barton:2003}
C.~H. Barton and A.~Sudbery, \textsl{ {Magic squares and matrix models of Lie
  algebras}},
\newblock Adv. in Math. \textbf{ 180}(2), 596--647 (2003), {math/0203010}.

\bibitem[CCGR13]{Carrasco:2012ca}
J.~J.~M. Carrasco, M.~Chiodaroli, M.~G{\"u}naydin and R.~Roiban, \textsl{
  {One-loop four-point amplitudes in pure and matter-coupled N = 4
  supergravity}},
\newblock JHEP \textbf{ 1303}, 056 (2013), {1212.1146}.

\bibitem[CCM15]{Cacciatori:2012cb}
S.~L. Cacciatori, B.~L. Cerchiai and A.~Marrani, \textsl{ {Squaring the
  Magic}},
\newblock Adv. Theor. Math. Phys. \textbf{ 19}, 923--954 (2015), {1208.6153}.

\bibitem[CGJR15a]{Chiodaroli:2014xia}
M.~Chiodaroli, M.~G{\"u}naydin, H.~Johansson and R.~Roiban, \textsl{
  {Scattering amplitudes in $ \mathcal{N}=2 $ Maxwell-Einstein and
  Yang-Mills/Einstein supergravity}},
\newblock JHEP \textbf{ 01}, 081 (2015), {1408.0764}.

\bibitem[CGJR15b]{Chiodaroli:2015rdg}
M.~Chiodaroli, M.~Gunaydin, H.~Johansson and R.~Roiban, \textsl{ {Spontaneously
  Broken Yang-Mills-Einstein Supergravities as Double Copies}},
\newblock (2015), {1511.01740}.

\bibitem[CGJR16]{Chiodaroli:2015wal}
M.~Chiodaroli, M.~Gunaydin, H.~Johansson and R.~Roiban, \textsl{ {Complete
  construction of magical, symmetric and homogeneous N=2 supergravities as
  double copies of gauge theories}},
\newblock Phys. Rev. Lett. \textbf{ 117}(1), 011603 (2016), {1512.09130}.

\bibitem[CGJR17]{Chiodaroli:2017ehv}
M.~Chiodaroli, M.~Gunaydin, H.~Johansson and R.~Roiban, \textsl{ {Gauged
  supergravities and spontaneous SUSY breaking from the double copy}},
\newblock (2017), {1710.08796}.

\bibitem[CGPT17]{Carrillo-Gonzalez:2017iyj}
M.~Carrillo-Gonzalez, R.~Penco and M.~Trodden, \textsl{ {The classical double
  copy in maximally symmetric spacetimes}},
\newblock (2017), {1711.01296}.

\bibitem[CGR12]{Chiodaroli:2011pp}
M.~Chiodaroli, M.~Gunaydin and R.~Roiban, \textsl{ {Superconformal symmetry and
  maximal supergravity in various dimensions}},
\newblock JHEP \textbf{ 1203}, 093 (2012), {1108.3085}.

\bibitem[Chi16]{Chiodaroli:2016jqw}
M.~Chiodaroli,
\newblock {Simplifying amplitudes in Maxwell-Einstein and Yang-Mills-Einstein
  supergravities},
\newblock 2016.

\bibitem[CHY14]{Cachazo:2013iea}
F.~Cachazo, S.~He and E.~Y. Yuan, \textsl{ {Scattering of Massless Particles:
  Scalars, Gluons and Gravitons}},
\newblock JHEP \textbf{ 1407}, 033 (2014), {1309.0885}.

\bibitem[CNN16]{Cardoso:2016ngt}
G.~L. Cardoso, S.~Nagy and S.~Nampuri, \textsl{ {A double copy for $
  \mathcal{N}=2 $ supergravity: a linearised tale told on-shell}},
\newblock JHEP \textbf{ 10}, 127 (2016), {1609.05022}.

\bibitem[CNN17]{Cardoso:2016amd}
G.~Cardoso, S.~Nagy and S.~Nampuri, \textsl{ {Multi-centered $ \mathcal{N}=2 $
  BPS black holes: a double copy description}},
\newblock JHEP \textbf{ 04}, 037 (2017), {1611.04409}.

\bibitem[DSW17]{DeSmet:2017rve}
P.-J. De~Smet and C.~D. White, \textsl{ {Extended solutions for the biadjoint
  scalar field}},
\newblock Phys. Lett. \textbf{ B775}, 163--167 (2017), {1708.01103}.

\bibitem[FM08]{Ferrara:2008de}
S.~Ferrara and A.~Marrani,
\newblock {Symmetric Spaces in Supergravity},
\newblock in \textsl{ Symmetry in Mathematics and Physics}, edited by V.~C.
  Donald~Babbitt and R.~Fioresi, volume 490 of \textsl{ Contemporary
  Mathematics}, pages 203--229, Providence, Rhode Island, 2008, American
  Mathematical Society,
\newblock Conference in honor of V.S. Varadarajan's 70th Birthday, Los Angeles,
  California, 18-20 Jan 2008.

\bibitem[Fre54]{Freudenthal:1954}
H.~Freudenthal, \textsl{ {Beziehungen der $E_7$ und $E_8$ zur oktavenebene
  I-II}},
\newblock Nederl. Akad. Wetensch. Proc. Ser. \textbf{ 57}, 218--230 (1954).

\bibitem[Fre59]{Freudenthal:1959}
H.~Freudenthal, \textsl{ {Beziehungen der $E_7$ und $E_8$ zur oktavenebene
  IX}},
\newblock Nederl. Akad. Wetensch. Proc. Ser. \textbf{ A62}, 466--474 (1959).

\bibitem[GL71]{Golfand:1971iw}
{\relax Yu}.~A. Golfand and E.~P. Likhtman, \textsl{ {Extension of the Algebra
  of Poincare Group Generators and Violation of p Invariance}},
\newblock JETP Lett. \textbf{ 13}, 323--326 (1971),
\newblock [Pisma Zh. Eksp. Teor. Fiz.13,452(1971)].

\bibitem[GNPW07]{Gunaydin:2007bg}
M.~Gunaydin, A.~Neitzke, B.~Pioline and A.~Waldron, \textsl{ {Quantum Attractor
  Flows}},
\newblock JHEP \textbf{ 09}, 056 (2007), {0707.0267}.

\bibitem[GP09]{Gunaydin:2009zza}
M.~Gunaydin and O.~Pavlyk, \textsl{ {Quasiconformal Realizations of $E(6)(6),
  E(7)(7), E(8)(8)$ and $SO(n+3,m+3)$, $N\geq4$ Supergravity and Spherical
  Vectors}},
\newblock Adv.Theor.Math.Phys. \textbf{ 13} (2009), {0904.0784}.

\bibitem[GPT17]{Goldberger:2017frp}
W.~D. Goldberger, S.~G. Prabhu and J.~O. Thompson, \textsl{ {Classical gluon
  and graviton radiation from the bi-adjoint scalar double copy}},
\newblock (2017), {1705.09263}.

\bibitem[GR16]{Goldberger:2016iau}
W.~D. Goldberger and A.~K. Ridgway, \textsl{ {Radiation and the classical
  double copy for color charges}},
\newblock (2016), {1611.03493}.

\bibitem[GST83]{Gunaydin:1983rk}
M.~G{\"u}naydin, G.~Sierra and P.~K. Townsend, \textsl{ {Exceptional
  supergravity theories and the magic square}},
\newblock Phys. Lett. \textbf{ B133}, 72 (1983).

\bibitem[GST84]{Gunaydin:1983bi}
M.~G{\"u}naydin, G.~Sierra and P.~K. Townsend, \textsl{ {The geometry of $N=2$
  Maxwell-Einstein supergravity and Jordan algebras}},
\newblock Nucl. Phys. \textbf{ B242}, 244 (1984).

\bibitem[GST85]{Gunaydin:1984ak}
M.~G{\"u}naydin, G.~Sierra and P.~K. Townsend, \textsl{ {Gauging the $d = 5$
  Maxwell-Einstein supergravity theories: More on Jordan algebras}},
\newblock Nucl. Phys. \textbf{ B253}, 573 (1985).

\bibitem[Gun93]{Gunaydin:1992zh}
M.~Gunaydin, \textsl{ {Generalized conformal and superconformal group actions
  and Jordan algebras}},
\newblock Mod.Phys.Lett. \textbf{ A8}, 1407--1416 (1993), {hep-th/9301050}.

\bibitem[HJ13]{Huang:2012wr}
Y.-t. Huang and H.~Johansson, \textsl{ {Equivalent D=3 Supergravity Amplitudes
  from Double Copies of Three-Algebra and Two-Algebra Gauge Theories}},
\newblock Phys.Rev.Lett. \textbf{ 110}, 171601 (2013), {1210.2255}.

\bibitem[HLL17a]{Henneaux:2016opm}
M.~Henneaux, V.~Lekeu and A.~Leonard, \textsl{ {Chiral Tensors of Mixed Young
  Symmetry}},
\newblock Phys. Rev. \textbf{ D95}(8), 084040 (2017), {1612.02772}.

\bibitem[HLL17b]{Henneaux:2017xsb}
M.~Henneaux, V.~Lekeu and A.~Leonard, \textsl{ {The Action of the (free)
  $(4,0)$-theory}},
\newblock (2017), {1711.07448}.

\bibitem[Hod13]{Hodges:2011wm}
A.~Hodges, \textsl{ {New expressions for gravitational scattering amplitudes}},
\newblock Journal of High Energy Physics \textbf{ 1307} (2013), {1108.2227}.

\bibitem[Hul00a]{Hull:2000zn}
C.~Hull, \textsl{ {Strongly coupled gravity and duality}},
\newblock Nucl.Phys. \textbf{ B583}, 237--259 (2000), {hep-th/0004195}.

\bibitem[Hul00b]{Hull:2000rr}
C.~Hull, \textsl{ {Symmetries and compactifications of (4,0) conformal
  gravity}},
\newblock JHEP \textbf{ 0012}, 007 (2000), {hep-th/0011215}.

\bibitem[Hur98]{Hurwitz:1898}
A.~Hurwitz, \textsl{ {Uber die komposition der quadratishen formen von beliebig
  vielen variabeln}},
\newblock Nachr. Ges. Wiss. Gottingen , 309--316 (1898).

\bibitem[Jac71]{Jacobson:1971}
N.~Jacobson,
\newblock \textsl{ Exceptional Lie algebras},
\newblock Marcel Dekker, Inc., New York, 1971.

\bibitem[JKM17]{Johansson:2017bfl}
H.~Johansson, G.~K{\"a}lin and G.~Mogull, \textsl{ {Two-loop supersymmetric QCD
  and half-maximal supergravity amplitudes}},
\newblock (2017), {1706.09381}.

\bibitem[JO15]{Johansson:2014zca}
H.~Johansson and A.~Ochirov, \textsl{ {Pure Gravities via Color-Kinematics
  Duality for Fundamental Matter}},
\newblock JHEP \textbf{ 11}, 046 (2015), {1407.4772}.

\bibitem[Jul80]{Julia:1980gr}
B.~Julia,
\newblock {Group disintegrations},
\newblock in \textsl{ Superspace and Supergravity}, edited by S.~Hawking and
  M.~Rocek, volume C8006162 of \textsl{ Nuffield Gravity Workshop}, pages
  331--350, Cambridge University Press, 1980.

\bibitem[JvNW]{Jordan:1933vh}
P.~Jordan, J.~von Neumann and E.~P. Wigner,
\newblock {On an algebraic generalization of the quantum mechanical formalism},
\newblock \href{http://www.jstor.org/stable/pdfplus/1968117.pdf}{\textit{Ann.
  Math.} \textbf{35} (1934) no. 1, 29--64}.

\bibitem[KT83]{Kugo:1982bn}
T.~Kugo and P.~K. Townsend, \textsl{ {Supersymmetry and the division
  algebras}},
\newblock Nucl. Phys. \textbf{ B221}, 357 (1983).

\bibitem[LM02]{Landsberg200259}
J.~Landsberg and L.~Manivel, \textsl{ Triality, Exceptional Lie Algebras and
  Deligne Dimension Formulas},
\newblock Advances in Mathematics \textbf{ 171}(1), 59 -- 85 (2002).

\bibitem[LM06]{landsberg2006sextonions}
J.~M. Landsberg and L.~Manivel, \textsl{ The sextonions and},
\newblock Advances in Mathematics \textbf{ 201}(1), 143--179 (2006).

\bibitem[LMN{\etalchar{+}}16]{Luna:2016due}
A.~Luna, R.~Monteiro, I.~Nicholson, D.~O'Connell and C.~D. White, \textsl{ {The
  double copy: Bremsstrahlung and accelerating black holes}},
\newblock JHEP \textbf{ 06}, 023 (2016), {1603.05737}.

\bibitem[LMN{\etalchar{+}}17]{Luna:2016hge}
A.~Luna, R.~Monteiro, I.~Nicholson, A.~Ochirov, D.~O'Connell, N.~Westerberg and
  C.~D. White, \textsl{ {Perturbative spacetimes from Yang-Mills theory}},
\newblock JHEP \textbf{ 04}, 069 (2017), {1611.07508}.

\bibitem[LMOW15]{Luna:2015paa}
A.~Luna, R.~Monteiro, D.~O'Connell and C.~D. White, \textsl{ {The classical
  double copy for Taub--NUT spacetime}},
\newblock Phys. Lett. \textbf{ B750}, 272--277 (2015), {1507.01869}.

\bibitem[LNOW17]{Luna:2017dtq}
A.~Luna, I.~Nicholson, D.~O'Connell and C.~D. White, \textsl{ {Inelastic Black
  Hole Scattering from Charged Scalar Amplitudes}},
\newblock (2017), {1711.03901}.

\bibitem[MO11]{Monteiro:2011pc}
R.~Monteiro and D.~O'Connell, \textsl{ {The Kinematic Algebra From the
  Self-Dual Sector}},
\newblock JHEP \textbf{ 1107}, 007 (2011), {1105.2565}.

\bibitem[MO14]{Monteiro:2013rya}
R.~Monteiro and D.~O'Connell, \textsl{ {The Kinematic Algebras from the
  Scattering Equations}},
\newblock JHEP \textbf{ 1403}, 110 (2014), {1311.1151}.

\bibitem[MOW14]{Monteiro:2014cda}
R.~Monteiro, D.~O'Connell and C.~D. White, \textsl{ {Black holes and the double
  copy}},
\newblock JHEP \textbf{ 1412}, 056 (2014), {1410.0239}.

\bibitem[MQS{\etalchar{+}}12]{Marrani:2012uu}
A.~Marrani, C.-X. Qiu, S.-Y.~D. Shih, A.~Tagliaferro and B.~Zumino, \textsl{
  {Freudenthal Gauge Theory}},
\newblock (2012), {1208.0013}.

\bibitem[Nag16]{Nagy:2014jza}
S.~Nagy, \textsl{ {Chiral Squaring}},
\newblock JHEP \textbf{ 07}, 142 (2016), {1412.4750}.

\bibitem[Ram71]{Ramond:1971gb}
P.~Ramond, \textsl{ {Dual Theory for Free Fermions}},
\newblock Phys. Rev. \textbf{ D3}, 2415--2418 (1971).

\bibitem[Ram77]{Ramond:1976aw}
P.~Ramond,
\newblock {Introduction to Exceptional Lie Groups and Algebras},
\newblock
  \href{http://ccdb4fs.kek.jp/cgi-bin/img/allpdf?197703037}{CALT-68-577}, 1977.

\bibitem[Ros56]{Rosenfeld:1956}
B.~A. Rosenfeld, \textsl{ {Geometrical interpretation of the compact simple Lie
  groups of the class $E$ (Russian)}},
\newblock Dokl. Akad. Nauk. SSSR \textbf{ 106}, 600--603 (1956).

\bibitem[SS]{Salam:1989fm}
A.~Salam and E.~Sezgin, editors,
\newblock \textsl{ Supergravities in diverse dimensions}, volume 1-2,
\newblock Netherlands: North-Holland, Amsterdam, 1989, 1499 p. Singapore: World
  Scientific, Singapore, 1989, 1499 p.

\bibitem[Str87]{Strathdee:1986jr}
J.~A. Strathdee, \textsl{ Extended {Poincar\'e} supersymmery},
\newblock Int. J. Mod. Phys. \textbf{ A2}, 273 (1987).

\bibitem[{Str}00]{Streater:2000}
{Streater, R. F. and Wightman, A. S.}, \textsl{ {PCT, spin and statistics, and
  all that}},
\newblock Princeton, USA: Princeton Univ. Pr. , 207 p. (2000).

\bibitem[Sud84]{Sudbery:1984}
A.~Sudbery, \textsl{ {Division algebras, (pseudo)orthogonal groups, and
  spinors}},
\newblock J. Phys. \textbf{ A17}(5), 939--955 (1984).

\bibitem[Tit55]{Tits:1955}
J.~Tits, \textsl{ {Interpr\'{e}tation g\'{e}om\'{e}triques de groupes de Lie
  simples compacts de la classe $E$}},
\newblock M\'{e}m. Acad. Roy. Belg. Sci \textbf{ 29}, 3 (1955).

\bibitem[Tit66]{Tits:1966}
J.~Tits, \textsl{ Alg\'ebres alternatives, alg\'ebres de Jordan et alg\'ebres
  de Lie exceptionnelles},
\newblock Indag. Math. \textbf{ 28}, 223--237 (1966).

\bibitem[VA72]{Volkov:1972jx}
D.~V. Volkov and V.~P. Akulov, \textsl{ {Possible universal neutrino
  interaction}},
\newblock JETP Lett. \textbf{ 16}, 438--440 (1972),
\newblock [Pisma Zh. Eksp. Teor. Fiz.16,621(1972)].

\bibitem[Var04]{Varadarajan:2004}
V.~S. Varadarajan,
\newblock \textsl{ {Supersymmetry for mathematicians: an introduction}},
\newblock American Mathematical Society, 2004.

\bibitem[Vin66]{Vinberg:1966}
E.~B. Vinberg, \textsl{ A construction of exceptional simple Lie groups},
\newblock Tr. Semin. Vektorn. Tr. Semin. Vektorn. Tensorn. Anal. \textbf{
  13}(7-9) (1966).

\bibitem[WB92]{Wess:1992}
J.~Wess and J.~Bagger,
\newblock \textsl{ Supersymmetry and supergravity},
\newblock Princeton University Press, second edition, 1992.

\bibitem[Wes06]{westbury2006sextonions}
B.~W. Westbury, \textsl{ Sextonions and the magic square},
\newblock Journal of the London Mathematical Society \textbf{ 73}(2), 455--474
  (2006).

\bibitem[Whi16]{White:2016jzc}
C.~D. White, \textsl{ {Exact solutions for the biadjoint scalar field}},
\newblock Phys. Lett. \textbf{ B763}, 365--369 (2016), {1606.04724}.

\bibitem[WZ74]{WESS197439}
J.~Wess and B.~Zumino, \textsl{ Supergauge transformations in four dimensions},
\newblock Nuclear Physics B \textbf{ 70}(1), 39 -- 50 (1974).

\end{thebibliography}
%\bibliographystyle{hep}

\newcommand{\etalchar}[1]{$^{#1}$}

\end{document}